\documentclass[
    twocolumn,
    amsmath,amstex,amssymb,
    aps, pra,
    floatfix,
    longbibliography,
    reprint]{revtex4-2}

\usepackage{graphicx}
\usepackage{dcolumn}%
\usepackage{bm}
\usepackage{lipsum}
\usepackage{gensymb}
\usepackage{physics}
\usepackage{xcolor}

\usepackage{booktabs}
\usepackage{ulem}

\usepackage[utf8]{inputenc}
\usepackage[T1]{fontenc}

\usepackage{hyperref}
\hypersetup{
  colorlinks   = true, 
  urlcolor     = blue, 
  linkcolor    = red, 
  citecolor   = red 
}

\begin{document}

\newcommand{\MyTitle}{Entangling Superconducting Qubits via Energy-Selective Local Reservoirs}

\title{\MyTitle}

\author{Qihao Guo}%
\author{Botao Du}
\author{Ruichao Ma}%
 \email{maruichao@purdue.edu}

\affiliation{%
Department of Physics and Astronomy, Purdue University, West Lafayette, IN 47907, USA }%

\date{August 12, 2026}

\begin{abstract}

Engineered dissipation provides a powerful route to controlling and stabilizing quantum states in open systems. Superconducting circuits are particularly suited to this approach due to their tunable coupling to dissipative environments. Here we realize programmable local reservoirs for superconducting qubits through parametrically driven couplings to readout resonators, creating energy-selective incoherent pump and loss. Using coupled superconducting qubits, we autonomously stabilize entangled single-excitation states with fidelity up to 90.8\%. We probe the stabilization dynamics under varying initial conditions and bath parameters, and implement robust classical shadow estimation for accurate and scalable state characterization. Finally, we numerically study a configuration where the engineered pump and loss share a common dissipative mode, leading to reservoir-mediated interference and classically correlated steady states. Our results demonstrate a scalable and hardware-efficient framework for dissipative preparation and control of correlated many-body states in superconducting circuits.

\end{abstract}

\maketitle

\section{Introduction}

Generating and controlling entanglement in engineered quantum platforms is central to quantum information science. Advances in coherent control have enabled increasingly precise manipulation of quantum states, ranging from high-fidelity gate operations in digital quantum processors to analog control of interacting many-body systems through tailored Hamiltonian dynamics. In superconducting circuits, strong nonlinearities, flexible circuit design, and high tunability have enabled programmable interactions and coherent control of increasingly large qubit arrays, leading to major advances in quantum computing and quantum simulation \cite{Blais2021-bn, Krantz2019-zv, Carusotto2020-ct}.

At the same time, dissipation and decoherence remain fundamental challenges for controlling complex quantum systems. Rather than viewing coupling to the environment solely as a limitation, however, the strong and tunable environment coupling available in engineered platforms has enabled engineered dissipation to become a resource for quantum control \cite{Verstraete2009-pj,Kapit2017-qj,Harrington2022-hl}. By designing suitable couplings between the system and tailored environments, driven-dissipative dynamics autonomously steer the system toward target quantum states. This approach is particularly natural in photonic and superconducting platforms, where finite excitation lifetimes necessitate continuous replenishment and stabilization.

In superconducting circuits, engineered dissipation has enabled autonomous cooling and qubit reset by transferring excitations into cold dissipative modes \cite{Valenzuela2006-cb,Geerlings2013-kl,Magnard2018-qm,Zhou2021-dc}. Combined with coherent driving, dissipative protocols can also stabilize metastable excited states and prepare multilevel states with tunable effective temperatures \cite{Cao2025-la}. In the many-body regime, incoherent pumping has further been used to stabilize photonic Mott insulating states through the interplay of interactions and dissipation \cite{Ma2019-ee}.
Engineered dissipation can also generate coherent correlations. Single-qubit superposition states are prepared using cavity-assisted bath engineering \cite{Murch2012-jy}, and universal stabilization has been demonstrated for a parametrically coupled qubit \cite{Lu2017-uq}. Dissipative protocols have enabled entanglement stabilization of two-qubit Bell states \cite{Shankar2013-ep,Li2024-kl,Brown2022-sg}, as well as few-qubit states using energy-resolved pumping \cite{Hacohen-Gourgy2015-zc} or Raman processes \cite{Chen2025-hc}. Engineered dissipation has also enabled remote entanglement mediated by dissipative waveguide interactions \cite{Shah2024-rn} and stabilization of protected bosonic states for autonomous quantum error correction \cite{Grimm2020-mj, Gertler2021-jd}. These approaches have been realized with a variety of hardware implementations, including dedicated on-chip dissipative elements such as engineered filters ~\cite{thorbeck2024high,ding2025multipurpose}, quantum-circuit refrigerators~\cite{tan2017quantum}, and nonlinear mixing elements~\cite{Cao2025-la}.

In many of these experiments, the stabilized coherence and entanglement structure are predominantly determined by externally applied coherent drives, parametrically engineered couplings, or symmetry-selective dissipation. A complementary perspective is to use local dissipative processes that do not directly impose spatial or phase coherence, allowing the stabilized states to emerge from the interacting many-body spectrum through energy-selective injection and removal of excitations. Motivated by how particle and thermal reservoirs stabilize strongly correlated phases in condensed matter systems, such approaches aim to realize nonequilibrium steady states analogous to equilibrium phases at finite chemical potential \cite{Kapit2014-zn,Hafezi2015-mw,Lebreuilly2017-hu}. Therefore, developing programmable local reservoirs with controlled spectral selectivity offers a promising route toward dissipative preparation and stabilization of correlated many-body states for quantum simulation and the generation of entangled resource states.

\begin{figure}[!t]
    \includegraphics[width=0.8\columnwidth]{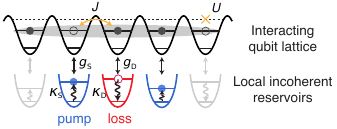}
    \caption{
    Interacting qubit lattice coupled to energy-selective local reservoirs. Transmon qubits realize a Bose-Hubbard lattice subject to local incoherent pump and loss processes. Our hardware-efficient implementation enables these local reservoirs to be dynamically controlled at each lattice site.
    }
    \label{fig:fig0_illus} 
\end{figure}

In this work, we implement hardware-efficient, energy-selective local reservoirs in superconducting circuits by utilizing the readout resonators of a standard flux-tunable transmon device as dissipative modes, without requiring dedicated hardware elements. The reservoir location, center frequency, and coupling strength can be dynamically programmed through the applied flux-modulation waveforms.
This is illustrated in Fig.\,\ref{fig:fig0_illus}, where a one-dimensional qubit lattice is coupled to local incoherent pump and loss processes with controlled spectral selectivity. We first discuss the implementation of parametrically engineered local reservoirs in Sec.\,\ref{sec:reservoir_eng}, and then demonstrate stabilization of entangled states in a coupled two-qubit system using local pumping and loss in Sec.\,\ref{sec:stabilization}. Finally, in Sec.\,\ref{sec:one_resonator} we investigate a scenario in which local pump and loss interact through a shared dissipative mode, leading to modified steady-state correlations.

\section{Qubit array with programmable local reservoirs}
\label{sec:reservoir_eng}

We consider a one-dimensional array of coupled transmon qubits with programmable local dissipation. In the strongly interacting limit, the system realizes a Bose-Hubbard model for microwave photons,
\begin{equation*}
 \mathcal{H}_{\text{BH}}/\hbar = \sum_{<ij>} J a_i^\dagger a_j + \frac{U}{2}\sum_i n_i(n_i-1) + \sum_i \epsilon_i n_i .
\end{equation*}
Here, $a_i$ is the bosonic annihilation operator on site $i$, $J$ is the nearest-neighbor tunneling rate, $n_i=a_i^\dagger a_i$ is the on-site occupation, $U$ is the on-site interaction, and $\epsilon_i$ is the site energy. Our experiments are performed on a device comprising four frequency-tunable transmons, each with individual frequency control and readout; a device image and detailed device parameters are presented in Refs.~\cite{Du2024-xp, Du2025-iw}. The qubits are tuned to a typical frequency of $\omega_q \approx 2\pi\times 4.5\,\mathrm{GHz}$ to form a degenerate lattice, with $J \approx 2\pi \times 6\,\mathrm{MHz}$ and $U\approx 2\pi \times -250\,\mathrm{MHz}$. 
Here we work in the hard-core limit with $J\ll |U|$ and $n_i \leq 1$, where the system is equivalently described by an $XY$ spin model~\cite{cazalilla2011one,yan2019strongly,Karamlou2024-gt}. The transmons have a typical relaxation rate $\Gamma_1 = 1/T_1 \approx 2\pi\times 5\,\mathrm{kHz}$ and local dephasing rate $\Gamma_\phi = 1/T_2^* \approx 2\pi\times 60\,\mathrm{kHz}$.

We engineer programmable local reservoirs that incoherently add or remove excitations from the qubit array. These energy-selective reservoirs are realized through parametric sideband interactions between the transmons and their readout resonators, induced by frequency modulation of the tunable transmons.
The modulation amplitude and frequency set the effective coupling strength $g_{\text{D/S}}$ and the detuning $\delta_{\text{D/S}}$ between the reservoir and the qubit array. We retain the notation $S$ and $D$ for the pump and loss, respectively, following the source-drain terminology used when treating microwave excitations as particles in the interacting transmon lattice.

The Hamiltonian of a single transmon lattice site coupled to its readout resonator is
\begin{equation*}
\mathcal{H}/\hbar = \omega_q(t)a^\dagger a + \omega_r b^\dagger b + g(a^\dagger + a)(b^\dagger + b) + \frac{U}{2}n(n-1),
\end{equation*}
where $b$ is the resonator annihilation operator. The qubit and resonator are capacitively coupled with strength $g$ and detuning $\Delta = \omega_r-\omega_q \gg g$. We apply qubit frequency modulation, $\omega_q(t) = \omega_q^0 + A_\text{mod} \cos(\omega_\text{mod} t)$, using the local flux-bias line.

For $\omega_\text{mod} \approx \omega_r-\omega_q^0$ (red sideband), the modulation induces resonant exchange between the qubit and the lossy readout resonator. The qubit excitation tunnels to the resonator and subsequently decays at a rate equal to the resonator linewidth $\kappa_r$. 
Hence, the red-sideband modulation implements a local, narrow-band loss.
In the rotating frame of the qubit and within the rotating-wave approximation, the effective Hamiltonian is
\begin{equation*}
\mathcal{H}_\text{D}/\hbar \approx \delta_\text{D} b^\dagger b + g_\text{D}(a^\dagger b + a b^\dagger) + \frac{U}{2}n(n-1),
\end{equation*}
with effective coupling
$g_\text{D} = g\, J_1(\frac{A_\text{mod}}{\omega_\text{mod}})$,
where $J_1$ is the first-order Bessel function of the first kind, and detuning
$\delta_\text{D} = (\omega_r - \omega_\text{mod}) - \omega_q^0$.
We work in the weak-coupling limit $g_\text{D} \ll \kappa_r$, where the reservoir bandwidth is set by the resonator linewidth, $\kappa_r \approx 2\pi\times 1.5\,\mathrm{MHz}$, through its coupling to the readout transmission line. In this limit, the qubit population decays at a detuning-dependent rate
$
\Gamma_\text{D}(\delta_\text{D})=\frac{g_\text{D}^2\,\kappa_r}{\delta_\text{D}^2+(\kappa_r/2)^2}
$,
which reduces on resonance to $\Gamma_\text{D}(0)=4g_\text{D}^2/\kappa_r$.

For $\omega_\text{mod} \approx \omega_r+\omega_q^0$ (blue sideband), the modulation generates a coherent two-photon process that creates or annihilates one excitation in the qubit and one in the resonator. 
Because the resonator photon is rapidly lost at rate $\kappa_r$, the blue sideband implements a narrow-band incoherent pump for the qubit array.
The effective Hamiltonian is
\begin{equation*}
\mathcal{H}_\text{S}/\hbar \approx -\delta_\text{S} b^\dagger b + g_\text{S}(a^\dagger b^\dagger + a b) + \frac{U}{2}n(n-1),
\end{equation*}
with effective coupling
$g_\text{S} = g\, J_1\left( \frac{A_\text{mod}}{\omega_\text{mod}}\right)$,
and detuning
$\delta_\text{S} = (\omega_\text{mod} - \omega_r) - \omega_q^0$.
In the same weak-coupling limit, the incoherent pump populates the qubit at the detuning-dependent rate
$\Gamma_\text{S}(\delta_\text{S})=\frac{g_\text{S}^2\,\kappa_r}{\delta_\text{S}^2+(\kappa_r/2)^2}$.
A detailed description of the reservoir implementation can be found in our previous work\cite{Du2024-xp, Du2025-iw}.

\section{Two-qubit Bell-state stabilization}
\label{sec:stabilization}

We now apply the energy-selective local reservoirs to stabilize entangled two-qubit states, focusing on single-excitation Bell states $\ket{\pm} = (\ket{ge} \pm \ket{eg})/\sqrt{2}$.

A typical experiment starts with two neighboring qubits far detuned (detuning $\Delta\gg J$) and prepared in a well-controlled initial state. The qubits are then tuned to resonance via fast flux control, and the flux modulations are turned on to enable the driven-dissipative reservoirs. After a variable interaction time, we turn off the flux modulations, and the two qubits are rapidly detuned to freeze population exchange. Subsequently, the two-qubit state is measured either directly in the population basis via multiplexed readout or, after applying single-qubit rotations, in arbitrary measurement bases. Additional details of the experimental sequences and the calibration of single-qubit gates and readout are discussed in Appendices \ref{sec:appendix_expt-seq} and \ref{sec:appendix_readout}.

While quantum state tomography (QST) can fully characterize the two-qubit driven-dissipative dynamics through reconstruction of the density matrix, the exponential cost of QST measurements makes it impractical to scale to larger systems. Classical shadow estimation \cite{Huang2020-cd} provides an alternative approach in which a large number of observables can be predicted efficiently using relatively few measurements. Superconducting circuit experiments are particularly well suited for shadow techniques due to the available local control, high-fidelity readout, and fast experimental cycles. However, standard shadow estimation is susceptible to systematic errors arising from noisy or imperfect unitaries and measurements. Recently, robust shadow estimation has been proposed \cite{Chen2021-cg} as a noise-resilient method for learning quantum state properties. In it, a calibration step is introduced in which noisy shadow measurements are performed on a high-fidelity initial state to learn the effective noise channel and subsequently mitigate its effects during the shadow-estimation procedure.
Here, we implement robust shadow estimation using the experimental procedures developed in \cite{Vermersch2024-tv, Vitale2024-oa} and demonstrate its efficacy in our two-qubit experiments.

\subsection{State preparation with incoherent pump}

\begin{figure}[!t]
    \includegraphics[width=1\columnwidth]{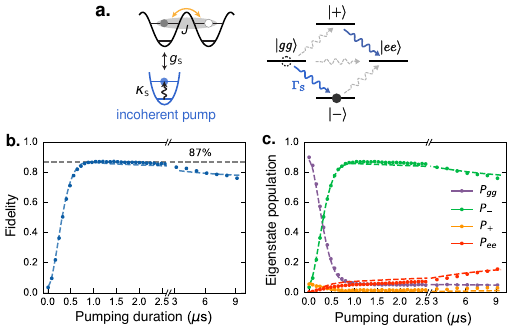}
    \caption{
Two-qubit dissipative stabilization.
(a) For two resonantly coupled qubits, a narrow-band incoherent pump selectively prepares a single-excitation eigenstate, as illustrated in the energy-level diagram. Solid arrows indicate transitions resonant with the pump for preparing $\ket{-}$, and dotted arrows indicate unwanted off-resonant transitions.
(b) Measured fidelity to $\ket{-}$ as a function of pump duration, starting from initial state $\ket{gg}$. 
(c) Measured eigenstate populations, showing leakage into $\ket{ee}$ at longer times. 
Dashed lines in (b,c) are numerical simulations using experimentally measured system parameters. Each experiment is typically averaged over \(90{,}000\) single-shot measurements. Fidelities and state populations are extracted using robust shadow estimation, with statistical error bars smaller than the marker size. See Appendices~\ref{sec:appendix_readout} and \ref{sec:appendix_uncertainty} for details of the state characterization and uncertainty estimation.
}
    \label{fig:2Q_pump} 
\end{figure}

For two transmon qubits resonantly coupled with tunneling $J$, $\ket{\pm}$ are the single-excitation eigenstates of the system with eigenenergies $\omega_\pm = \omega_q \pm J$, as shown in Fig.~\ref{fig:2Q_pump}(a). 
When the energy selectivity of the incoherent pump is much narrower than the eigenenergy splitting ($\kappa_r \ll 2J$, as in our experiments), the system can be incoherently pumped from the ground state $\ket{gg}$ into individual eigenstates.
Experimentally, we initialize two neighboring qubits $Q_1$ and $Q_2$ in the ground state. We then turn on flux modulation on $Q_1$ to enable the incoherent pump with detuning $\delta_{\text{S}} \approx -J$, flux-modulation-induced coupling $g_\text{S} = 2\pi \times 0.75~\text{MHz}$, and corresponding on-resonance pumping rate $\Gamma_{\text{S}} = 4g^2_\text{S}/\kappa_r \approx 2\pi\times 1.5~\text{MHz}$. 

In Fig.~\ref{fig:2Q_pump}(b), we plot the measured fidelity as a function of pump duration, obtained using robust shadow estimation. The fidelity between the instantaneous state with density matrix $\rho$ and a target pure state $\ket{\psi_{\mathrm{tar}}}$ is defined as $F(\rho,\ket{\psi_{\mathrm{tar}}})=\bra{\psi_{\mathrm{tar}}}\rho\ket{\psi_{\mathrm{tar}}}$. We reach a maximum fidelity of approximately $87\%$ with $\ket{\psi_{\mathrm{tar}}}=\ket{-}$ at about $1\,\mu\text{s}$, consistent with the timescale set by the engineered pump rate, $\tau = 2\pi/\Gamma_\text{S}$.
At longer pump times, the fidelity gradually decreases due to unwanted excitation by the pump and decoherence processes. Because the bath has a finite linewidth $\kappa_r$, it also induces off-resonant transitions. In the present case, the dominant leakage process is pumping from $|-\rangle$ to $|ee\rangle$, which is detuned by $2J$ from the pump resonance. In the weak-coupling limit, this occurs at the off-resonant rate
$\Gamma_\text{S}(\delta_\text{S}=2J)=\frac{g_\text{S}^2\,\kappa_r}{(2J)^2+(\kappa_r/2)^2}
\approx 2\pi\times 6\,\mathrm{kHz}$.
Once populated, the $|ee\rangle$ state relaxes at the intrinsic qubit decay rate $\Gamma_1$ into the single-excitation manifold and eventually toward $\ket{gg}$.
This competition between off-resonant incoherent pumping and intrinsic decay results in a finite incoherent population in $|ee\rangle$ over time. This is observed in Fig.~\ref{fig:2Q_pump}(c), where we plot the eigenstate populations extracted from density matrices obtained via quantum state tomography. The incoherent pump can also induce the transition from $|gg\rangle$ to $|ee\rangle$ via a two-photon process with intermediate-state detuning $J$, but this process is much slower in our experiments. For all data presented, we observe good agreement between experiments and numerical simulations. The detailed numerical model is presented in Appendix \ref{sec:appendix_modeling}.

\begin{figure}[!t]
    \includegraphics[width=1.0\columnwidth]{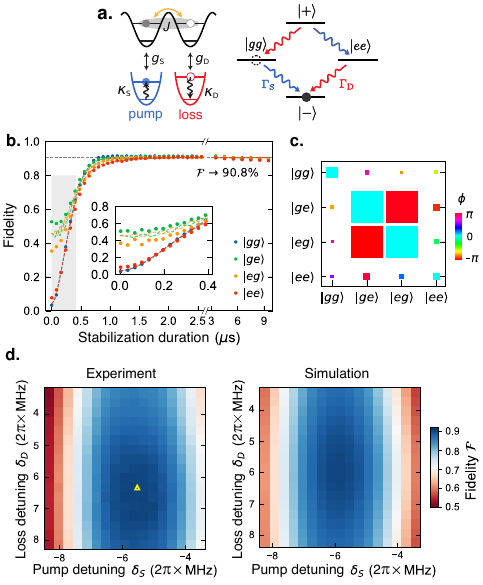}
    \caption{
    Stabilization with pump and loss.
    (a) Adding a narrow-band loss on the second qubit selectively removes leakage population and stabilizes the target state.
    (b) Measured stabilization dynamics from different initial Fock states, showing convergence to a steady state with \(90.8\%\) fidelity to \(\ket{-}\) within \(2\,\mu\mathrm{s}\). Inset: zoom-in of the short-time dynamics. Dashed lines are numerical simulations using experimentally measured parameters. Statistical error bars are smaller than the marker size.
    (c) Representative stabilized density matrix measured at \(t=2.8\,\mu\mathrm{s}\).
    (d) Fidelity versus pump and loss detunings at a stabilization duration of \(3\,\mu\mathrm{s}\), measured experimentally (left) and compared with numerical simulation (right). The bath coupling rates are fixed at \(g_{\mathrm{S}} = g_{\mathrm{D}} = 2\pi\times 0.73~\mathrm{MHz}\). The triangular marker indicates the detunings used in (b,c).
    }   
    \label{fig:2Q_pumpcool}
\end{figure}

\subsection{Stabilization with pump and loss}

We now add the energy-selective loss and show that it continuously removes leakage population and stabilizes the desired state, here $\ket{-}$. The loss reservoir is applied at detuning $\delta_{\mathrm{D}} \approx +J$ with sideband coupling rate $g_{\mathrm{D}} = 2\pi \times 0.58~\mathrm{MHz}$ on $Q_2$, concurrent with the incoherent pump applied on $Q_1$, as illustrated in Fig.~\ref{fig:2Q_pumpcool}(a).
In Fig.~\ref{fig:2Q_pumpcool}(b), we show the resulting stabilization dynamics toward $\ket{-}$. After approximately $2~\mu\mathrm{s}$, the system reaches its steady state with a fidelity of $F=90.8\%$ to $\ket{-}$. A representative density matrix measured at $t = 2.8~\mu\mathrm{s}$ is shown in Fig.~\ref{fig:2Q_pumpcool}(c).
To show that the stabilization is robust to the initial state, we measure the stabilization dynamics starting from different two-site initial states $\{\ket{gg}, \ket{ge}, \ket{eg}, \ket{ee}\}$ and observe convergence to the target state, with the steady-state fidelity maintained for more than $10~\mu\mathrm{s}$.
In Appendix~\ref{sec:appendix_2Q-concurrence}, we further characterize the entanglement of the stabilized state using the measured concurrence~\cite{Wootters1998-rd}, which reaches a steady-state value of $C\approx 0.82$.

The fidelity of the dissipatively stabilized $\ket{-}$ state is relatively insensitive to small changes in the pump and loss detunings, provided they remain small or comparable to the bath bandwidth.
In Fig.\,~\ref{fig:2Q_pumpcool}(d), we show the fidelity as a function of the pump and loss detunings, $\delta_{\text{S}}$ and $\delta_{\text{D}}$, at a stabilization duration of \(3\,\mu\mathrm{s}\). The stabilization duration is chosen to be sufficiently long to reach the steady state near the optimal detunings. The experimental data are in good agreement with numerical modeling, where the offset in optimal detuning is attributed to a small miscalibration of flux-modulation-induced ac-Stark shifts on the qubits.
Similarly, the measured optimal fidelity remains relatively insensitive to changes in bath coupling rates; see discussion below on the sources of infidelity.

\begin{figure}[]
    \includegraphics[width=0.65\columnwidth]{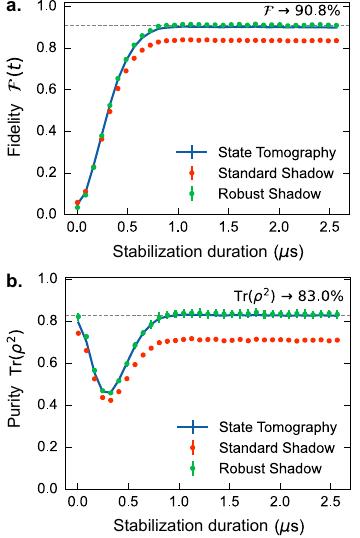}
    \caption{
    Comparison of quantum state tomography, standard classical shadow, and robust shadow estimation.
    (a) Fidelity to the $\ket{-}$ state for the stabilization experiment in Fig.\,\ref{fig:2Q_pumpcool}, starting from the initial state $\ket{gg}$. Robust shadow estimation agrees closely with quantum state tomography, while standard classical shadow estimation exhibits systematic deviations.
    (b) State purity as a function of stabilization duration. See Appendix \ref{sec:appendix_uncertainty} for details on data analysis and error estimation.
    }
    \label{fig:SI_shadow-comparison} 
\end{figure}

In Fig.\,\ref{fig:SI_shadow-comparison}, we compare the results obtained from stabilizing the two-qubit system starting from the $|gg\rangle$ state, using three different characterization methods: quantum state tomography, standard classical shadow, and robust shadow.
The quantum state tomography data are reconstructed after careful calibration of all systematic errors of our single-qubit rotations and readout. Standard classical shadow, which assumes perfect unitary and measurement, shows significant deviations in fidelity and purity during the dynamics.
In contrast, robust shadow estimation mitigates these errors and gives results consistent with quantum state tomography. This demonstrates the effectiveness of robust shadow estimation for efficient and flexible state characterization in future experiments involving larger qubit arrays. In Appendix \ref{sec:appendix_rshadow}, we summarize the experimental and analysis protocols for robust shadow estimation, along with an uncertainty analysis based on our measurement parameters. 

The same stabilization scheme can stabilize the $\ket{+}$ state by reversing the pump and loss detunings $\delta_{\text{S}/\text{D}}$. In Appendix \ref{sec:appendix_2Q-extra}, we show additional data for stabilizing $\ket{+}$ with a fidelity of approximately $88.5\%$. The slightly lower steady-state fidelity for the $\ket{+}$ state, compared to that of $\ket{-}$ ($F=90.8\%$), is due to the higher intrinsic relaxation rate of $\ket{+}$, with a lifetime of $19\,\mu\mathrm{s}$ compared to $50\,\mu\mathrm{s}$ for $\ket{-}$. In the absence of sideband interactions, transmon relaxation at the lattice frequencies is partially limited by intrinsic Purcell decay through the resonators. The $\ket{+}$ ($\ket{-}$) state therefore exhibits an enhanced (suppressed) decay rate due to constructive (destructive) interference of emission from the two qubits into the common readout transmission line.

\subsection{Sources of stabilization infidelity}

\begin{figure*}[!th]
    \includegraphics[width=1.5\columnwidth]{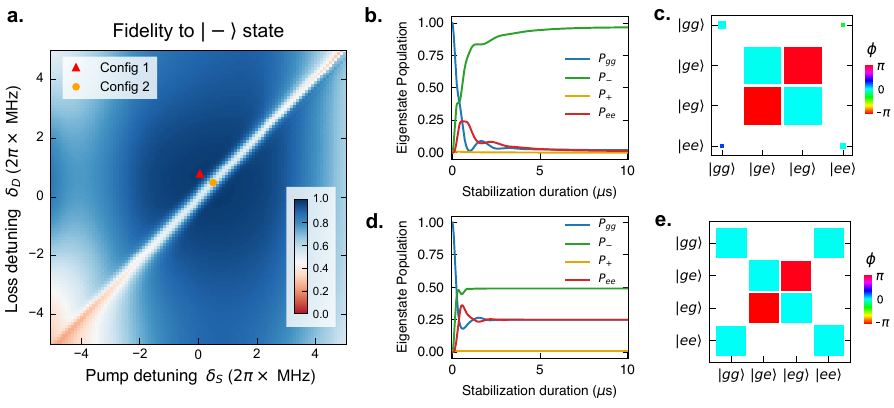}
    \caption{
    Interference effects from incoherent pump and loss implemented through a shared dissipative mode. Results are obtained from numerical simulation.
    (a) Steady-state fidelity to \(|-\rangle\) versus pump and loss detunings \(\delta_{\rm S}\) and \(\delta_{\rm D}\) applied simultaneously to the same qubit.
    (b,c) Eigenstate population dynamics and steady-state density matrix for mismatched detunings (configuration 1), showing stabilization of \(|-\rangle\).
    (d,e) For matched detunings (configuration 2), a Raman-like process stabilizes the separable \(X\)-state \(\rho_{\rm X}\) with coherences in both the even- and odd-parity manifolds.
    }
    \label{fig:SI_2Q-single-res} 
\end{figure*}

We now discuss the main mechanisms that limit the observed steady-state fidelity to $\ket{-}$, supported by numerical simulations. 
The dominant source of infidelity is thermal population in the engineered reservoirs. We measure an average thermal population of $2.5\%$ in the readout resonators. For the loss reservoir, the finite effective temperature of the resonator can resonantly drive thermal excitations into the qubit array, resulting in unwanted heating. Numerical simulations indicate that the measured resonator thermal population contributes approximately $5.5\%$ infidelity to stabilization of $\ket{-}$.

A second limitation arises from intrinsic lattice relaxation, which competes directly with the engineered stabilization dynamics. We independently measure $T_1^{\ket{+}} \approx 19\,\mu\mathrm{s}$, $T_1^{\ket{-}} \approx 50\,\mu\mathrm{s}$, and $T_1^{\ket{ee}} \approx 12\,\mu\mathrm{s}$. The longer lifetime of $\ket{-}$ compared to $\ket{+}$ is attributed to interference between the two qubits' Purcell decay channels through the readout resonators. For the symmetric wavefunction $\ket{+}$, the emitted amplitudes from the two qubits interfere constructively in the common readout transmission line, leading to enhanced decay. In addition, we perform a Ramsey experiment between the $\ket{\pm}$ states and measure a dephasing time of $13\,\mu\mathrm{s}$, indicating that decoherence in this manifold primarily arises from relaxation toward $\ket{gg}$. Pure dephasing within the single-excitation manifold is strongly suppressed by the large resonant tunneling $J$ \cite{Levine2024-yd_truncate}.
Numerical simulations attribute approximately $2.4\%$ infidelity in $\ket{-}$ stabilization to intrinsic decoherence under our experimental conditions. The shorter (longer) lifetime of $\ket{+}$ ($\ket{-}$) is also consistent with the lower (higher) observed stabilization fidelity for $\ket{+}$ ($\ket{-}$).

Lastly, the finite linewidth of the reservoirs leads to off-resonant driving of unwanted transitions. This produces incoherent mixtures whose populations are determined by the relative strengths of the frequency-dependent pump and loss rates \cite{Cao2025-la}. 
In the present case, this results in a residual incoherent population in $\ket{+}$ that scales as $(\kappa_r/2J)^2 \approx 1.6\%$. Although this contribution can be reduced by improving spectral selectivity, doing so must be balanced against the need for sufficiently fast pumping and loss rates, which set the overall preparation and stabilization timescales.
Recently, it has been demonstrated that the trade-off between dissipative preparation time and fidelity can be circumvented via dark-state engineering utilizing multilevel interactions \cite{Brown2022-sg}.

The observed infidelity is well accounted for by the three error channels discussed above and analyzed in detail in Appendix~\ref{sec:appendix_infidelity}, all of which can be substantially suppressed using established experimental techniques. Improved thermalization and filtering of the cryogenic microwave environment ~\cite{yeh2017microwave,PhysRevApplied.11.014031,krinner2019engineering} can reduce the resonator thermal population to below $0.1\%$~\cite{Zhou2021-dc}, at which level our numerical simulations predict an improved steady-state fidelity of approximately $96\%$; the fidelity is insensitive to the residual qubit thermal population.
Intrinsic decoherence, comprising single-qubit decay and two-qubit collective decay, currently contributes an infidelity of $2.4\%$; this contribution can be reduced through state-of-the-art transmon design and fabrication~\cite{place2021new,wang2022towards}, together with on-chip filtering to suppress Purcell decay through the readout circuitry~\cite{PhysRevLett.112.190504}.
The off-resonant sideband transitions can be mitigated by improving the spectral selectivity of the drives. Simulations indicate that reducing the ratio to $\kappa_r/J \approx 1/8$, while keeping $g_{\text{D/S}}\sim \kappa_r$ to maintain efficient stabilization, yields fidelities exceeding $99\%$.

\section{Interference effects in\\ a shared reservoir}
\label{sec:one_resonator}

Finally, we explore a scenario in which the incoherent pump and loss are implemented using a single dissipative mode, i.e., the flux modulations used to engineer the two reservoirs are applied simultaneously to the same qubit.

When both reservoirs are applied through a shared lossy mode, the sideband transitions can coherently interfere through common intermediate states. In the two-qubit case, the red and blue sidebands combine to induce an effective Raman-like two-photon process connecting \(|gg\rangle\) and \(|ee\rangle\) through the single-excitation manifold. This process generates coherence within the even-parity manifold, while the original dissipative stabilization processes maintain coherence in the odd-parity manifold. For reservoir detunings that stabilize the antisymmetric state \(|-\rangle\) in the regular two-mode implementation (\(\delta_{\rm S/D}\approx \mp J\)), balanced pump and loss rates applied through the same dissipative mode lead to a steady state approximately described by
\begin{equation*}
\rho_{\rm X}
=
\frac{1}{2}|-\rangle\!\langle-|
+
\frac{1}{2}|\Phi^+\rangle\!\langle\Phi^+|,
\qquad
|\Phi^+\rangle
=
\frac{|gg\rangle+|ee\rangle}{\sqrt{2}} .
\end{equation*}
In the computational basis, this state is
\begin{equation*}
\rho_{\rm X}
=
\frac{1}{4}
\begin{pmatrix}
1 & 0 & 0 & 1 \\
0 & 1 & -1 & 0 \\
0 & -1 & 1 & 0 \\
1 & 0 & 0 & 1
\end{pmatrix},
\end{equation*}
which has nonzero elements in an \(X\)-shaped pattern.
Despite being an equal mixture of two maximally entangled Bell states, \(\rho_{\rm X}\) is separable. It admits the decomposition
\begin{equation*}
\rho_{\rm X}
=
\frac{1}{2}|+_y,-_y\rangle\!\langle+_y,-_y|
+
\frac{1}{2}|-_y,+_y\rangle\!\langle-_y,+_y|,
\end{equation*}
where
$|\pm_y\rangle = (|g\rangle\pm \mathrm{i}|e\rangle)/\sqrt{2}$.
The steady state therefore exhibits perfect classical correlations in the transverse basis, despite containing no entanglement, with
\begin{equation*}
\langle Y_1 Y_2\rangle=-1,
\qquad
\langle X_1 X_2\rangle
=
\langle Z_1 Z_2\rangle
=
0.
\end{equation*}
When the pump and loss detunings are mismatched (\(|\delta_{\rm S}|\neq|\delta_{\rm D}|\)), the two-photon Raman process connecting \(|gg\rangle\) and \(|ee\rangle\) becomes off-resonant and the even-parity coherence is suppressed. In this regime, the shared dissipative mode recovers the same dissipative stabilization of the \(|-\rangle\) state obtained in the standard implementation using separate pump and loss reservoirs.

Figure~\ref{fig:SI_2Q-single-res}(a) shows the numerically simulated steady-state fidelity to \(|-\rangle\) as a function of the pump and loss detunings applied to the same qubit. High fidelity to \(|-\rangle\) is observed over a broad range of detunings except near \(\delta_{\rm S}\approx\delta_{\rm D}\), where the fidelity decreases to approximately \(50\%\). Representative eigenstate population dynamics and steady states for mismatched and matched detunings are shown in Fig.~\ref{fig:SI_2Q-single-res}(b)--(e), including stabilization of the expected \(X\)-state in Fig.~\ref{fig:SI_2Q-single-res}(d),(e). Stabilizing the $X$-state relies on coherent interference between the red- and blue-sideband processes and therefore requires a stable, calibrated phase relation between the corresponding flux-modulation pulses. Although unavailable in our present control setup, this phase coherence can be implemented in future experiments using phase-coherent microwave generation.

This example illustrates a distinct regime of dissipative control in which interference between engineered pump and loss stabilizes basis-selective coherence and strong transverse correlations without generating entanglement. More broadly, it provides a controlled setting to probe the boundary between classical and quantum correlations in open systems, and suggests a route toward engineering correlated but separable states in larger many-body architectures.

\section{Outlook}

Our hardware-efficient implementation of local dissipative reservoirs provides a scalable route to stabilizing many-body states in superconducting lattices. Because incoherent pump and loss are implemented using standard readout resonators with flux-modulated coupling, the reservoirs can be dynamically activated at arbitrary lattice sites with minimal hardware overhead and integrated with coherent control, opening routes to tailoring correlations and entanglement in interacting quantum systems \cite{Chu2025-vx, Guo2025-rj}.

A central question in larger systems is which classes of many-body phases can be efficiently stabilized through local dissipative reservoirs \cite{Mi2024-in_truncate, Zhan2025-xx}. Narrow-band, eigenstate-selective reservoirs, as demonstrated here, are well suited to few-body systems where the level spacing exceeds the reservoir bandwidth. In larger lattices, the many-body level spacing decreases with system size, intrinsic decoherence becomes an increasingly broadband decay channel, and stabilization from arbitrary initial states requires reservoirs that connect many eigenstates simultaneously. These considerations naturally point toward engineered pump and loss reservoirs with tailored broadband spectra that act as an effective chemical potential for interacting photons, enabling dissipative preparation of correlated ground-state-like phases \cite{Marcos2012-ep, Hafezi2015-mw, Lebreuilly2016-py, Lebreuilly2017-hu}. Efficient pumping and cooling further depend on the spatial structure of the reservoir coupling \cite{Ma2017-vw, Kurilovich2022-zj}. Controlled spectral overlap between the reservoirs can also impose effective detailed balance, stabilizing thermal many-body ensembles with a tunable effective temperature, relevant to finite-temperature quantum simulation and quantum Gibbs-sampling algorithms \cite{Shabani2016-go, Cao2025-la}.

Beyond steady-state preparation, engineered reservoirs enable controlled studies of nonequilibrium quantum dynamics and dissipative many-body phases \cite{Lewis-Swan2019-qd, Yanay2020-go}. Their microscopic dynamics provide insight into information propagation and thermalization in driven quantum systems, including the role of initial conditions in phenomena such as quantum Mpemba effects \cite{Ares2025-qf}. These directions further connect to emerging efforts in quantum thermodynamics with superconducting circuits \cite{Campbell2025-il_truncate, Aamir2025-kz, Uusnakki2026-ll}.

\section*{Acknowledgments}

The authors thank Senrui Chen for discussions and assistance with implementing robust classical shadows, and Qi Zhou for discussions on dissipative stabilization.
This work was supported by grants from the National Science Foundation (award number DMR-2145323) and the Air Force Office of Scientific Research (award number FA9550-23-1-0491). 

\section*{Data availability}

The data presented in this work, together with the numerical simulation and analysis code used to generate the results, are openly available at Ref.\,\cite{guo_2026_20190733}.

\appendix

\section{Experimental sequence}
\label{sec:appendix_expt-seq}


A typical experiment consists of state initialization, driven-dissipative evolution under the engineered reservoirs, and final state measurement.

At the start of each sequence, all qubits are tuned to the same idle frequency using DC flux biases. Fast flux pulses are applied throughout the experiment to dynamically detune the qubit frequencies for initialization and readout. In thermal equilibrium, each qubit has an excited-state population of approximately $5\%$. To suppress errors in state preparation and readout calibration arising from residual qubit thermal population, we perform a measurement-based heralding step at the beginning of each experimental sequence. The qubits are first detuned by approximately $80\,\text{MHz}$ and measured in the computational basis. The measurement outcomes are then used to postselect the true ground state $\ket{gg}$, reducing the residual excited-state population to below approximately $0.8\%$. After heralding, we wait $1.4\,\mu\text{s}$ for residual readout photons to decay before proceeding with the experiment.

For experiments requiring product-state initialization, such as Fig.\,\ref{fig:2Q_pumpcool}(b) and readout calibration, the qubits are detuned and individually addressed with resonant single-qubit pulses applied through the readout transmission line. The single-qubit gates have a typical duration of $40$\,ns and Gaussian envelopes with $\sigma = 8$\,ns. After initialization, the qubits are rapidly tuned back to a degenerate frequency, and flux modulation is applied through the individual flux-bias lines to activate the engineered incoherent pump and/or loss for a variable evolution time.

At the end of the reservoir-coupled evolution, the flux modulation is turned off, and the qubits are rapidly moved to their readout frequencies, where neighboring qubits are detuned by at least $400$\,MHz to suppress residual interactions during measurement. Measurements in bases other than the computational basis are implemented by sequentially applying single-qubit rotations before readout. The qubit states are measured using frequency-multiplexed dispersive readout with a typical pulse duration of $1.7\,\mu$s. After readout, flux-balancing pulses are applied to each flux line to cancel the net accumulated current during the sequence and reduce slow qubit-frequency drifts.

The experiment is repeated with a cycle period of $300\,\mu$s, allowing the qubits and resonators to relax back to thermal equilibrium between experiments. Each experiment presented in the paper is typically repeated $90{,}000$ times, acquired as 9 independent runs of $10{,}000$ shots each.

\section{Readout characterization}
\label{sec:appendix_readout}

To improve readout fidelity, we implement a shelving readout scheme \cite{Elder2020-ya}. All experiments in the paper are performed in the two-level limit, with negligible population outside the transmon states $\ket{g}$ and $\ket{e}$. Before readout, a resonant $\pi_{\mathrm{ef}}$ pulse transfers any population in $\ket{e}$ to $\ket{f}$. Shelving the excited-state population suppresses readout errors arising from qubit relaxation during the measurement pulse.

To calibrate the readout discriminator, we initialize the system in all computational basis states and record the corresponding single-shot multiplexed $I/Q$ signals. A linear support vector machine (SVM) is then trained for state discrimination. The calibration states are prepared using sequential $\pi$ pulses in the readout configuration, with neighboring qubits far detuned to suppress drive crosstalk. The combined contribution of pulse errors and $T_1$ decay during the sequential preparation results in a state-preparation error of approximately $0.6\%$ for two-qubit Fock states. Using the shelving protocol, we achieve average single-qubit $\ket{g}$-$\ket{e}$ assignment fidelity of $98.4\%$.

For direct population measurements and quantum state tomography, readout-error mitigation is performed using the measured assignment probability matrix. Ensemble-averaged outcome probabilities obtained from the SVM are multiplied by the inverse assignment matrix to reconstruct the underlying Fock-state populations.
For both standard and robust classical shadow estimation, the SVM-predicted single-shot bit strings are used directly. In the robust shadow estimation, readout-error mitigation is incorporated directly into the shadow reconstruction procedure.

\textit{Phase calibration for single-qubit rotations:}
When the qubits are moved from the degenerate frequencies to the detuned readout configurations, each qubit accumulates a fixed dynamical phase that must be calibrated to implement the correct single-qubit rotations. We determine this phase $\phi$ by preparing a state with a known relative phase between the qubits. 

For the two-qubit experiments, this is achieved by coherently driving a resonant $\pi$ pulse on the symmetric $\ket{+}$ state \cite{Du2024-xp}. Quantum state tomography is then performed without compensating for the dynamical phase. The accumulated phase $\phi$ is extracted either from the reconstructed density matrix or by maximizing the correlator $\langle \sigma^x_1 \sigma^x_2 + \sigma^y_1 \sigma^y_2 \rangle$.
For the detunings used in this work, the accumulated dynamical phase is typically $\sim 2.3$ radians and remains stable across experiments.
\section{Calibration of qubit and resonator thermal populations}
\label{sec:appendix_calibration}

Here we describe the calibration procedure for the thermal populations of the qubits and the readout resonators.

\begin{figure*}[!th]
    \includegraphics[width=1.5\columnwidth]{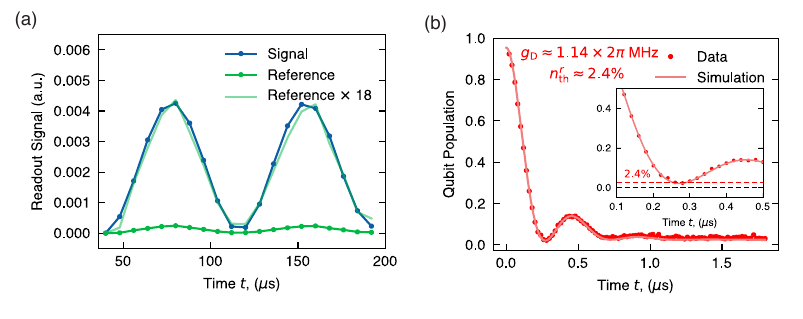}
    \caption{Calibration of qubit and resonator thermal populations.
(a) Qubit thermal population measured via $|e\rangle-|f\rangle$ Rabi oscillations. Rabi oscillations on the $|e\rangle-|f\rangle$ transition are recorded with a preceding $|g\rangle-|e\rangle~\pi$ pulse (Signal, blue) and without it (Reference, green). The reference oscillation, arising only from the residual thermal population in $|e\rangle$, is smaller by a factor of $\sim 18$, corresponding to an equilibrium excited-state population of approximately $5\%$. (b) Resonator thermal population measured via resonant sideband swaps. The qubit is initialized in $|e\rangle$ and undergoes damped vacuum-Rabi-like oscillations with its readout resonator under a strong, resonant red-sideband interaction ($g_{\text{D}}\approx 2\pi\times 1.14$~MHz). Red points: measured qubit population; solid line: master-equation simulation with the resonator thermal population $n_\text{th}^r$ and the effective sideband rate as fit parameters, yielding $n_\text{th}^r\approx2.4\%$. Inset: zoom-in of the first population minimum; the residual qubit population (red dashed line) lies above the zero-temperature expectation (black dashed line), reflecting the finite resonator thermal occupation.}
    \label{fig:SI_thermal_pop_calibration} 
\end{figure*}

\textit{\textbf{Qubit thermal population:}}
In thermal equilibrium, each transmon retains a residual excited-state population of approximately $5\%$, which affects both the stabilization experiments and the calibration of the readout reference signals. We calibrate this residual population using the $|e\rangle-|f\rangle$ Rabi technique of Ref.~\cite{jin2015thermal}, in which the amplitudes of Rabi oscillations on the $|e\rangle-|f\rangle$ transition, measured with and without a preceding $|g\rangle-|e\rangle~\pi$ pulse (see Fig.~\ref{fig:SI_thermal_pop_calibration}(a)), are compared to extract the equilibrium excited-state population. The calibrated thermal population is then used to correct the readout reference signals: because the nominally prepared basis states contain finite thermal populations, they correspond to known mixtures of bare Fock states rather than pure states. We therefore apply a linear transformation, constructed from the calibrated thermal populations, that converts populations measured with respect to the prepared basis states into populations in the bare Fock basis.

\textit{\textbf{Resonator thermal population:}} We measure the resonator thermal population by swapping resonator excitations onto the qubit using a strong, resonant sideband interaction, with a swap time shorter than the resonator ringdown time $1/\kappa_r$. For the incoherent loss, we initialize the qubit in its first excited state and record its population under the resonant red-sideband interaction as a function of the interaction time (Fig.~\ref{fig:SI_thermal_pop_calibration}(b)). The measured dynamics are fit to master-equation simulations performed in QuTiP, with the resonator thermal population and the effective sideband rate as the fit parameters. For the incoherent pump, the same procedure is performed starting from the qubit in its thermal-equilibrium ground state and applying a strong resonant blue-sideband interaction. From these measurements, we extract a thermal population of $2.4-
3\%$ for the readout resonators.

\section{Numerical modeling}
\label{sec:appendix_modeling}

For the numerical simulations, the dynamics of the qubit lattice coupled to energy-selective incoherent pump and loss reservoirs are modeled using a Lindblad master equation:
\begin{equation*}
\frac{d\rho}{dt}
= -\frac{\mathrm{i}}{\hbar}[H,\rho]
+ \sum_{m} \mathcal{D}[L_m]\rho,
\end{equation*}
where each incoherent process is described by a Lindblad jump operator $L_m$:
\begin{equation}
\mathcal{D}[L_m]\rho
= L_m\rho L_m^\dagger - \frac{1}{2}\left\{L_m^\dagger L_m,\rho\right\}.
\end{equation}

The Hamiltonian $H$ includes both the Bose-Hubbard lattice realized by the transmon array and the effective resonator-lattice coupling induced by the sideband flux modulation described in the main text. As a concrete two-qubit example, we consider $Q_1$ coupled to an incoherent pump reservoir via a blue-sideband interaction with resonator $R_1$, and $Q_2$ coupled to an incoherent loss reservoir via a red-sideband interaction with resonator $R_2$. Truncating the qubits to two levels and applying the rotating-wave approximation gives the effective Hamiltonian
\begin{equation}
\begin{aligned}
H/\hbar = & \underbrace{J(a^\dagger_1 a_2 + a_1 a^\dagger_2)}_{\text{qubit lattice}}
+\underbrace{\delta_{\text{S}} b^\dagger_1 b_1 + g_{\text{S}}(a^\dagger_1 b^\dagger_1 + a_1 b_1)}_{\text{pump reservoir}} \\
& - \underbrace{\delta_{\text{D}} b^\dagger_2 b_2 + g_{\text{D}}(a^\dagger_2 b_2 + a_2 b^\dagger_2)}_{\text{loss reservoir}}\,.
\end{aligned}
\end{equation}

The Lindblad model includes resonator decay (with linewidth $\kappa_{r_i}$) as the source of dissipation generating the effective incoherent reservoirs; qubit relaxation including both intrinsic single-qubit decay ($\Gamma_1$) and the experimentally measured collective Purcell decay rates $\Gamma_{+/-}$; and finite thermal populations in both the qubits ($n^q_{\text{th}}$) and resonators ($n^r_{\text{th}}$), although the qubit thermal population does not affect the stabilized steady state. Due to the strong coherent interaction $J$, we observe no significant pure dephasing within the $|\pm\rangle$ manifold and therefore neglect pure dephasing in the numerical model.
The full set of jump operators $L_m$ included in the model is:
\begin{itemize}
\item Pump-reservoir dissipation and thermal excitation: \\
    $\sqrt{\kappa_{r_1}(1+n^r_{\text{th}})}\, b_1,\quad
    \sqrt{\kappa_{r_1}n^r_{\text{th}}}\, b^\dagger_1$\,.
    \item Loss-reservoir dissipation and thermal excitation: \\
    $\sqrt{\kappa_{r_2}(1+n^r_{\text{th}})}\, b_2,\quad
    \sqrt{\kappa_{r_2}n^r_{\text{th}}}\, b^\dagger_2$\,.
    
    \item Single-qubit relaxation and thermal excitation: \\
    $\sqrt{\Gamma_1 (1+n^q_{\text{th}})}\, a_n,\quad
    \sqrt{\Gamma_1 n^q_{\text{th}}}\, a^\dagger_n,\qquad n=1,2$\,.
    
    \item Collective decay and thermal excitation: \\
    $\sqrt{\Gamma_{\pm}(1+n^q_{\text{th}})}\, |gg\rangle \langle \pm|,\quad
    \sqrt{\Gamma_{\pm}n^q_{\text{th}}}\, |\pm\rangle \langle gg|$\,,
    $\sqrt{\Gamma_{\pm}(1+n^q_{\text{th}}})\, |\pm\rangle \langle ee|,\quad
    \sqrt{\Gamma_{\pm}n^q_{\text{th}}}\, |ee\rangle \langle \pm|$\,.
    
\end{itemize}

Parameters used for the simulation results shown in Fig.\,\ref{fig:2Q_pump} and Fig.\,\ref{fig:2Q_pumpcool} are all independently measured: $J = 2\pi\times 6.0$ MHz. Coupling to the incoherent pump reservoir $g_{\text{S}} = 2\pi\times  0.75$ MHz, and coupling to the loss reservoir $g_{\text{D}} = 2\pi\times  0.58$ MHz.
Resonator linewidth $\kappa_{r_i} = 2\pi\times 1.5$ MHz. Intrinsic single-qubit decay $\Gamma_1 = 2\pi\times 4.0$\,kHz, measured collective Purcell decay rates $\Gamma_{+/-} = 2\pi\times  8.4/3.2$\,kHz for the $|\pm\rangle$ states~\cite{du2026programmable}. Qubit thermal populations $n^q_{\text{th}} = 5.0\%$ and resonator thermal populations $n^r_{\text{th}} = 2.5\%$.

\section{Analysis of Infidelity Sources}
\label{sec:appendix_infidelity}
\begin{figure*}[!th]
    \includegraphics[width=2.0\columnwidth]{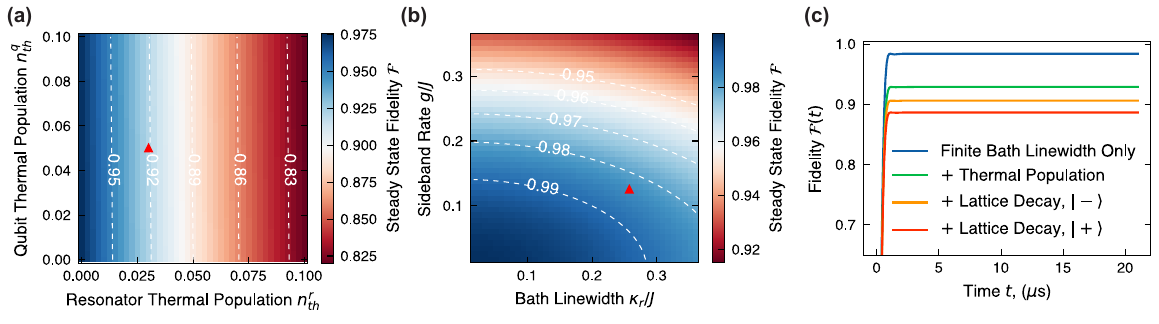}
    \caption{Numerical analysis of stabilization imperfections. (a) Steady-state fidelity $\mathcal{F}$ as a function of the resonator and qubit thermal populations, $n^q_{\text{th}}$ and $n^r_{\text{th}}$. White dashed lines are contours of constant fidelity; the red triangle marks the parameters calibrated in our experiment. The nearly vertical contours indicate that the fidelity is limited predominantly by the resonator thermal population. (b) Steady-state fidelity as a function of the bath linewidth $\kappa_r/J$ and the sideband rate $g/J$. The red triangle marks the experimental operating point. (c) Fidelity dynamics $\mathcal{F}$ under successively added infidelity sources: finite bath linewidth only (blue), with resonator thermal population (green), and with lattice decay with qubit thermal population for the target states $|-\rangle$ (orange) and $|+\rangle$ (red).}
    \label{fig:SI_infidelity_sources}
\end{figure*}

To identify the dominant error mechanisms, we numerically simulate the steady-state fidelity as a function of the system imperfections. Fig.~\ref{fig:SI_infidelity_sources} (a) shows the fidelity versus the thermal populations of the resonator and qubit. The contours of constant fidelity are nearly vertical, indicating that the fidelity is far more sensitive to the resonator thermal population than to that of the qubit; at our calibrated operating point (red triangle, the parameters in Appendix~\ref{sec:appendix_modeling}), the thermal occupation of the resonator is the leading contribution to infidelity. Fig.~\ref{fig:SI_infidelity_sources}(b) shows the fidelity, computed in the absence of resonator thermal population and the two-qubit decay, in the plane of bath linewidth $\kappa_r/J$ and the pump sideband rate $g/J$. The map reveals the trade-off between spectral selectivity and stabilization rate: fidelities exceeding $99\%$ are attainable by modestly reducing $\kappa_r/J$ while keeping $g\sim \kappa_r$ to maximize the stabilization efficiency, consistent with the discussion in the main text. Finally, Fig.~\ref{fig:SI_infidelity_sources}(c) evaluates the fidelity dynamics at the calibrated parameters of the present experiment, corresponding to the red triangles in Figs.~\ref{fig:SI_infidelity_sources}(a) and (b). Adding the infidelity sources one at a time decomposes the total infidelity into contributions from the finite bath linewidth, the thermal population, and the lattice decay of the target states $|\pm\rangle$.

\section{Classical shadow estimation}
\label{sec:appendix_rshadow}

Classical shadow estimation provides an efficient and scalable method for estimating expectation values $\langle O\rangle = \mathrm{Tr}(\hat{O}\rho)$ of observables $\hat{O}$ for an unknown $N$-qubit quantum state $\rho$. Here, we summarize the standard classical shadow estimation \cite{Huang2020-cd} and the robust shadow estimation that mitigates systematic errors arising from imperfect gates and measurements \cite{Chen2021-cg}. For the latter, we follow the modified implementation developed in Refs.\,\cite{Vermersch2024-tv, Vitale2024-oa}.

\textit{\textbf{Standard shadow estimation:}}

\textbf{Step 1}: \textit{Apply random local unitaries.} For an unknown quantum state $\rho$, we apply a random local Clifford unitary $U_i = U_{i,0} \otimes U_{i,1} \otimes \cdots \otimes U_{i,N-1}$, where each single-qubit operation is chosen from $U_{i,n} \in \{X_{\pi/2}, Y_{\pi/2}, \mathrm{Id}\}$ ($n = 0,\dots,N-1$). The resulting state $U_i \rho U_i^\dagger$ is then measured in the computational basis $\{|\vec{b}\rangle = |b_0,b_1,\cdots,b_{N-1}\rangle,\ b_n \in \{g,e\}\}$.

\textbf{Step 2}: \textit{Construct classical snapshots.} For the $i$-th experimental shot ($N_S$ shots in total, with $N_S = N_U N_M$), the classical shadow snapshot is reconstructed as
\begin{equation}\label{EqS.2.1}
\rho_i = \bigotimes_{n=0}^{N-1} \left(3U^\dagger_{i,n} |b_n\rangle \langle b_n| U_{i,n} - \mathbb{I} \right)\,.
\end{equation}
Averaging over random Clifford unitaries $\{U_i\}$ with measurement outcomes sampled according to the conditional probability $P(\vec{b}|U_i)$, the expected value of the reconstructed snapshot satisfies $\mathbb{E}_{U_i, \vec{b}}[\rho_i] = \rho$,
thereby enabling reconstruction of the original quantum state.

\textbf{Step 3}: \textit{Estimate observables using the median-of-means estimator.} The snapshots $\{\rho_i\}$ are divided into $K$ groups. For each group,
\begin{equation}\label{EqS.2.2}
\bar{O}^{(k)} = \frac{1}{N_S/K} \sum^{k N_S/K}_{i = (k-1)N_S/K + 1} \mathrm{Tr}(\hat{O}\rho_i)\,, \qquad k = 1,\dots, K\,,
\end{equation}
and the estimator for $\langle O\rangle$ is obtained from the median over the groups:
\begin{equation}\label{EqS.2.3}
\langle O\rangle = \mathrm{median} \{\bar{O}^{(1)}, \bar{O}^{(2)}, \dots ,\bar{O}^{(K)}\}\,.
\end{equation}

Classical shadows can also estimate nonlinear functions of the density matrix, such as the purity $\mathrm{Tr}(\rho^2)$. For each group,
\begin{equation}\label{EqS.2.4}
\mathrm{Tr}(\rho^2)^{(k)} = \frac{1}{N_S/K (N_S/K-1)} \sum_{i\neq j} \mathrm{Tr}(\rho_i\rho_j)
\end{equation}
where $i,j \in \{(k-1)N_S/K + 1, \dots, k N_S/K \}$ and $k = 1,\dots, K$. The final purity estimate is then
\begin{equation}\label{EqS.2.5}
\mathrm{Tr}(\rho^2) = \mathrm{median} \{\mathrm{Tr}(\rho^2)^{(1)}, \mathrm{Tr}(\rho^2)^{(2)}, \dots ,\mathrm{Tr}(\rho^2)^{(K)}\}\,.
\end{equation}

\textit{\textbf{Robust shadow estimation: }}

Here, the effective noise channel $\Lambda_n$ for each qubit $n$, arising from gate and measurement errors, is first characterized using a calibration experiment and then compensated through classical postprocessing.

Compared to the standard shadow estimation, the reconstruction of the snapshots is modified to be:
\begin{equation}\label{EqS.2.6}
\rho_i = \bigotimes_{n=0}^{N-1} \left(\alpha_n U^\dagger_{i,n} |b_n\rangle \langle b_n| U_{i,n} - \beta_n \mathbb{I} \right),
\end{equation}
where $\alpha_n = \frac{3}{2G_n - 1}$ and $\beta_n = \frac{2 - G_n}{2G_n - 1}$, with $G_n = \frac{1}{2} \langle b_n | \Lambda_n(|b_n\rangle \langle b_n|) | b_n \rangle$ representing the average survival probability of the two computational basis states under the noise channel $\Lambda_n$ for qubit $n$.

The calibration experiment begins from the high-fidelity ground state $|\vec{g}\rangle = |g\rangle^{\otimes N}$. Random local Clifford operations $U_i \in \{X_{\pi/2}, Y_{\pi/2}, \mathrm{Id}\}^{\otimes N}$ are applied, followed by measurement in the computational basis to obtain the conditional probabilities $P(\vec{b}|U_i)$. We then compute
\begin{equation}\label{EqS.2.7}
\begin{aligned}
C_n = \frac{1}{N_U} \sum_i \sum_{b_n = g,e} P(b_n | U_i)\,
|\langle b_n| U_{i,n} |g\rangle|^2\,,
\end{aligned}
\end{equation}
where $P(b_n | U_i)$ is the marginal probability of measuring outcome $b_n \in \{g,e\}$ for qubit $n$ after applying $U_i$.

The average survival probability for each qubit is then
\begin{equation}\label{EqS.2.8}
G_n = 3C_n - 1\,,
\end{equation}
from which the coefficients $\{\alpha_n,\beta_n\}$ are determined~\cite{Vitale2024-oa}. The robust shadow snapshots in Eq.\,\ref{EqS.2.6} are then used in place of the standard snapshots in Eq.\,\ref{EqS.2.1}, while observables are estimated using the same median-of-means procedure described above.

\begin{figure}[t]
    \includegraphics[width=0.98\columnwidth]{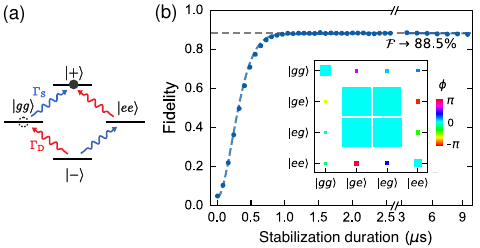}
    \caption{Stabilization of the $|+\rangle$ State.
(a) Illustration for the pump and loss detuning.
(b) Dynamics of the fidelity to the target state, when the initial state is $\ket{gg}$. The dashed line is the numerical simulation. Inset: Reconstructed density matrix of the stabilized steady-state at $t = 2.5\,\mu$s.}
    \label{fig:SI_2Q-plus-state} 
\end{figure}

\section{Data and uncertainty estimation}
\label{sec:appendix_uncertainty}

For readout calibration, we run 90,000 single-shot experiments for each computational basis state. After postselection on the heralding measurement result to effectively eliminate the thermal population, we retain about 50\% of the shots to train the SVM discriminator and obtain the readout assignment probability matrix. The statistical error in the assignment matrix is below 0.1\%.

For the stabilization experiments, each sequence is repeated $90{,}000$ times, acquired as 9 independent runs of $10{,}000$ shots each. This same dataset is used for the robust shadow estimation (Figs.\,\ref{fig:2Q_pump}, \ref{fig:2Q_pumpcool}, \ref{fig:SI_shadow-comparison}, and \ref{fig:SI_2Q-plus-state}), standard shadow estimation (Fig.\,\ref{fig:SI_shadow-comparison}), and quantum state tomography (Fig.\,\ref{fig:SI_shadow-comparison}). For robust shadow, the calibration data is also 90,000 shots.

For the classical shadow analysis, the fidelity and eigenstate populations are extracted by estimating the expectation values of projectors onto the states $\ket{gg}$, $\ket{\psi_+}$, $\ket{\psi_-}$, and $\ket{ee}$, where the single-excitation eigenstates $\ket{\psi_\pm} = \frac{1}{\sqrt{2}} \left( \ket{ge} \pm e^{-i\phi}\ket{eg} \right)$ include the calibrated dynamical phase $\phi$ (see Appendix~\ref{sec:appendix_readout}). The corresponding populations are given by $P_i = \langle \psi_i | \rho | \psi_i \rangle$, which also yields the fidelity to the target state. The density matrix is reconstructed by directly estimating the expectation values of the full set of two-qubit Pauli observables $[I,X,Y,Z]^{\otimes 2}$ using the shadow estimation procedure. The median-of-means estimator is performed by splitting all shadow snapshots into $K=100$ equally sized groups, calculating the expectation values for each group, and then taking the median of these values.

The error bars for the robust shadow estimation shown in the figures are obtained from bootstrap resampling of the measured shadow snapshots \cite{elben2025randommeasjl}. For each dataset, snapshots are resampled with replacement from the measured data, and the observable is re-estimated for up to 400 bootstrap realizations. The quoted uncertainty corresponds to the standard deviation of the resulting bootstrap distribution. This procedure is performed independently for each of the 9 experimental datasets, and the final uncertainty is obtained from the standard error of the mean across the independently bootstrapped datasets. The same procedure is used for the standard shadow analysis.

The average statistical error bars for the robust shadow estimates are 0.37\% for the state populations and fidelities, and $1.9$\% for the purity. The larger uncertainty for the purity arises because it depends on higher-order combinations of Pauli correlators. For the standard shadow analysis, the average statistical error bars are $0.18$\% for the state populations and fidelities, and $0.76$\% for the purity.
The statistical uncertainty of the robust shadow estimator is slightly larger than that of the standard shadow estimator because the calibration procedure introduces additional fluctuations associated with the independently measured calibration dataset. In contrast, for the standard shadow analysis, systematic errors arising from finite readout fidelity and imperfect single-qubit rotations dominate over the statistical uncertainty.

For the quantum state tomography (QST) analysis in Fig.\,\ref{fig:SI_shadow-comparison}, the expectation values of the two-qubit Pauli observables are measured directly, with readout-error mitigation performed using the measured readout assignment probability matrix. The density matrix is then reconstructed using maximum-likelihood estimation (MLE), which enforces a physical density matrix satisfying positivity and unit trace \cite{PhysRevLett.108.070502}.

The QST error bars represent the standard error of the mean estimated from the MLE-reconstructed density matrices of 9 independent datasets, with an average value of 0.14\%. The resulting statistical uncertainties are smaller than those obtained from the shadow estimation with comparable measurement statistics. This reduction arises because the MLE procedure constrains the reconstructed density matrix to remain within the physical state space, suppressing fluctuations of the reconstructed matrix elements at the cost of potentially introducing reconstruction bias.

\section{Stabilization of $\ket{+}$ state}
\label{sec:appendix_2Q-extra}

In the main text, we present data demonstrating the stabilization of the $|-\rangle$ state. In Fig.\,\ref{fig:SI_2Q-plus-state}, we show the stabilization of the $|+\rangle$ state by switching the source and drain detunings to $(\delta_{\text{S}}, \delta_{\text{D}}) = 2\pi\times (5.83, 6.23)$\,MHz. The sideband coupling rates are $g_\text{S} = g_\text{D} = 2\pi\times 0.73$\,MHz. We achieve a stabilization fidelity of $88.5\%$ for the $|+\rangle$ state. 

\section{Concurrence of the stabilized states}
\label{sec:appendix_2Q-concurrence}
\begin{figure}[t]
    \includegraphics[width=0.7\columnwidth]{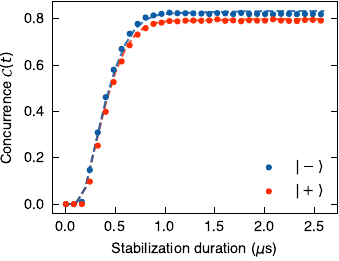}
    \caption{Entanglement of the stabilized states. Concurrence $C(t)$ as a function of stabilization duration for the stabilization of $|-\rangle$ (blue) and $|+\rangle$ (red), starting from the initial state $|gg\rangle$. Dashed lines show numerical simulations using experimental parameters.}
    \label{fig:SI_2Q-concurrence} 
\end{figure}

To directly quantify the entanglement of the dissipatively stabilized states, we compute the concurrence~\cite{Wootters1998-rd} from the density matrices reconstructed via robust shadow estimation. For a two-qubit state $\rho$, the concurrence is defined as $C(\rho) = \max(0,\lambda_1-\lambda_2-\lambda_3-\lambda_4)$, where $\lambda_i$ are the square roots of the eigenvalues, in decreasing order, of the matrix $\rho\tilde{\rho}$, with $\tilde{\rho} = (\sigma_y\otimes \sigma_y)\rho^* (\sigma_y\otimes \sigma_y)$ and the complex conjugate taken in the computational basis. The density matrices are reconstructed from the full set of two-qubit Pauli observables estimated with the robust shadow procedure described in Appendix~\ref{sec:appendix_rshadow}. We note that, unlike the fidelity and purity, the concurrence is a nonlinear function of the density matrix and is therefore computed from the reconstructed state rather than estimated directly from the shadow snapshots. Fig.~\ref{fig:SI_2Q-concurrence} shows the concurrence as a function of stabilization duration for the stabilization of both the $|-\rangle$ and $|+\rangle$ states, starting from the initial state $|gg\rangle$. The concurrence builds up on the stabilization timescale, in agreement with numerical simulations, and reaches steady-state values of $C\approx 0.82$ and $0.79$ for the $|-\rangle$ and $|+\rangle$ states, respectively, directly certifying the entanglement generated by the engineered local reservoirs. These values are consistent with the concurrence lower bound $C \geq 2F - 1$, which gives the minimum concurrence compatible with the measured Bell-state fidelity of the stabilized state.

\bibliography{paperpile, extra_ref_diss-stab}

\begin{thebibliography}{66}%
\makeatletter
\providecommand \@ifxundefined [1]{%
 \@ifx{#1\undefined}
}%
\providecommand \@ifnum [1]{%
 \ifnum #1\expandafter \@firstoftwo
 \else \expandafter \@secondoftwo
 \fi
}%
\providecommand \@ifx [1]{%
 \ifx #1\expandafter \@firstoftwo
 \else \expandafter \@secondoftwo
 \fi
}%
\providecommand \natexlab [1]{#1}%
\providecommand \enquote  [1]{``#1''}%
\providecommand \bibnamefont  [1]{#1}%
\providecommand \bibfnamefont [1]{#1}%
\providecommand \citenamefont [1]{#1}%
\providecommand \href@noop [0]{\@secondoftwo}%
\providecommand \href [0]{\begingroup \@sanitize@url \@href}%
\providecommand \@href[1]{\@@startlink{#1}\@@href}%
\providecommand \@@href[1]{\endgroup#1\@@endlink}%
\providecommand \@sanitize@url [0]{\catcode `\\12\catcode `\$12\catcode `\&12\catcode `\#12\catcode `\^12\catcode `\_12\catcode `\%12\relax}%
\providecommand \@@startlink[1]{}%
\providecommand \@@endlink[0]{}%
\providecommand \url  [0]{\begingroup\@sanitize@url \@url }%
\providecommand \@url [1]{\endgroup\@href {#1}{\urlprefix }}%
\providecommand \urlprefix  [0]{URL }%
\providecommand \Eprint [0]{\href }%
\providecommand \doibase [0]{https://doi.org/}%
\providecommand \selectlanguage [0]{\@gobble}%
\providecommand \bibinfo  [0]{\@secondoftwo}%
\providecommand \bibfield  [0]{\@secondoftwo}%
\providecommand \translation [1]{[#1]}%
\providecommand \BibitemOpen [0]{}%
\providecommand \bibitemStop [0]{}%
\providecommand \bibitemNoStop [0]{.\EOS\space}%
\providecommand \EOS [0]{\spacefactor3000\relax}%
\providecommand \BibitemShut  [1]{\csname bibitem#1\endcsname}%
\let\auto@bib@innerbib\@empty
\bibitem [{\citenamefont {Blais}\ \emph {et~al.}(2021)\citenamefont {Blais}, \citenamefont {Grimsmo}, \citenamefont {Girvin},\ and\ \citenamefont {Wallraff}}]{Blais2021-bn}%
  \BibitemOpen
  \bibfield  {author} {\bibinfo {author} {\bibfnamefont {A.}~\bibnamefont {Blais}}, \bibinfo {author} {\bibfnamefont {A.~L.}\ \bibnamefont {Grimsmo}}, \bibinfo {author} {\bibfnamefont {S.~M.}\ \bibnamefont {Girvin}},\ and\ \bibinfo {author} {\bibfnamefont {A.}~\bibnamefont {Wallraff}},\ }\bibfield  {title} {\bibinfo {title} {Circuit quantum electrodynamics},\ }\href {https://journals.aps.org/rmp/abstract/10.1103/RevModPhys.93.025005} {\bibfield  {journal} {\bibinfo  {journal} {Rev. Mod. Phys.}\ }\textbf {\bibinfo {volume} {93}} (\bibinfo {year} {2021})}\BibitemShut {NoStop}%
\bibitem [{\citenamefont {Krantz}\ \emph {et~al.}(2019)\citenamefont {Krantz}, \citenamefont {Kjaergaard}, \citenamefont {Yan}, \citenamefont {Orlando}, \citenamefont {Gustavsson},\ and\ \citenamefont {Oliver}}]{Krantz2019-zv}%
  \BibitemOpen
  \bibfield  {author} {\bibinfo {author} {\bibfnamefont {P.}~\bibnamefont {Krantz}}, \bibinfo {author} {\bibfnamefont {M.}~\bibnamefont {Kjaergaard}}, \bibinfo {author} {\bibfnamefont {F.}~\bibnamefont {Yan}}, \bibinfo {author} {\bibfnamefont {T.~P.}\ \bibnamefont {Orlando}}, \bibinfo {author} {\bibfnamefont {S.}~\bibnamefont {Gustavsson}},\ and\ \bibinfo {author} {\bibfnamefont {W.~D.}\ \bibnamefont {Oliver}},\ }\bibfield  {title} {\bibinfo {title} {A quantum engineer's guide to superconducting qubits},\ }\href {https://doi.org/10.1063/1.5089550} {\bibfield  {journal} {\bibinfo  {journal} {Appl. Phys. Rev.}\ }\textbf {\bibinfo {volume} {6}},\ \bibinfo {pages} {021318} (\bibinfo {year} {2019})}\BibitemShut {NoStop}%
\bibitem [{\citenamefont {Carusotto}\ \emph {et~al.}(2020)\citenamefont {Carusotto}, \citenamefont {Houck}, \citenamefont {Koll\'{a}r}, \citenamefont {Roushan}, \citenamefont {Schuster},\ and\ \citenamefont {Simon}}]{Carusotto2020-ct}%
  \BibitemOpen
  \bibfield  {author} {\bibinfo {author} {\bibfnamefont {I.}~\bibnamefont {Carusotto}}, \bibinfo {author} {\bibfnamefont {A.~A.}\ \bibnamefont {Houck}}, \bibinfo {author} {\bibfnamefont {A.~J.}\ \bibnamefont {Koll\'{a}r}}, \bibinfo {author} {\bibfnamefont {P.}~\bibnamefont {Roushan}}, \bibinfo {author} {\bibfnamefont {D.~I.}\ \bibnamefont {Schuster}},\ and\ \bibinfo {author} {\bibfnamefont {J.}~\bibnamefont {Simon}},\ }\bibfield  {title} {\bibinfo {title} {Photonic materials in circuit quantum electrodynamics},\ }\href {https://www.nature.com/articles/s41567-020-0815-y} {\bibfield  {journal} {\bibinfo  {journal} {Nat. Phys.}\ }\textbf {\bibinfo {volume} {16}},\ \bibinfo {pages} {268} (\bibinfo {year} {2020})}\BibitemShut {NoStop}%
\bibitem [{\citenamefont {Verstraete}\ \emph {et~al.}(2009)\citenamefont {Verstraete}, \citenamefont {Wolf},\ and\ \citenamefont {Cirac}}]{Verstraete2009-pj}%
  \BibitemOpen
  \bibfield  {author} {\bibinfo {author} {\bibfnamefont {F.}~\bibnamefont {Verstraete}}, \bibinfo {author} {\bibfnamefont {M.~M.}\ \bibnamefont {Wolf}},\ and\ \bibinfo {author} {\bibfnamefont {J.~I.}\ \bibnamefont {Cirac}},\ }\bibfield  {title} {\bibinfo {title} {Quantum computation and quantum-state engineering driven by dissipation},\ }\href {http://dx.doi.org/10.1038/nphys1342} {\bibfield  {journal} {\bibinfo  {journal} {Nat. Phys.}\ }\textbf {\bibinfo {volume} {5}},\ \bibinfo {pages} {633} (\bibinfo {year} {2009})}\BibitemShut {NoStop}%
\bibitem [{\citenamefont {Kapit}(2017)}]{Kapit2017-qj}%
  \BibitemOpen
  \bibfield  {author} {\bibinfo {author} {\bibfnamefont {E.}~\bibnamefont {Kapit}},\ }\bibfield  {title} {\bibinfo {title} {The upside of noise: engineered dissipation as a resource in superconducting circuits},\ }\href {https://iopscience.iop.org/article/10.1088/2058-9565/aa7e5d/meta} {\bibfield  {journal} {\bibinfo  {journal} {Quantum Sci. Technol.}\ }\textbf {\bibinfo {volume} {2}},\ \bibinfo {pages} {033002} (\bibinfo {year} {2017})}\BibitemShut {NoStop}%
\bibitem [{\citenamefont {Harrington}\ \emph {et~al.}(2022)\citenamefont {Harrington}, \citenamefont {Mueller},\ and\ \citenamefont {Murch}}]{Harrington2022-hl}%
  \BibitemOpen
  \bibfield  {author} {\bibinfo {author} {\bibfnamefont {P.~M.}\ \bibnamefont {Harrington}}, \bibinfo {author} {\bibfnamefont {E.~J.}\ \bibnamefont {Mueller}},\ and\ \bibinfo {author} {\bibfnamefont {K.~W.}\ \bibnamefont {Murch}},\ }\bibfield  {title} {\bibinfo {title} {Engineered dissipation for quantum information science},\ }\href {https://www.nature.com/articles/s42254-022-00494-8} {\bibfield  {journal} {\bibinfo  {journal} {Nature Reviews Physics}\ }\textbf {\bibinfo {volume} {4}},\ \bibinfo {pages} {660} (\bibinfo {year} {2022})}\BibitemShut {NoStop}%
\bibitem [{\citenamefont {Valenzuela}\ \emph {et~al.}(2006)\citenamefont {Valenzuela}, \citenamefont {Oliver}, \citenamefont {Berns}, \citenamefont {Berggren}, \citenamefont {Levitov},\ and\ \citenamefont {Orlando}}]{Valenzuela2006-cb}%
  \BibitemOpen
  \bibfield  {author} {\bibinfo {author} {\bibfnamefont {S.~O.}\ \bibnamefont {Valenzuela}}, \bibinfo {author} {\bibfnamefont {W.~D.}\ \bibnamefont {Oliver}}, \bibinfo {author} {\bibfnamefont {D.~M.}\ \bibnamefont {Berns}}, \bibinfo {author} {\bibfnamefont {K.~K.}\ \bibnamefont {Berggren}}, \bibinfo {author} {\bibfnamefont {L.~S.}\ \bibnamefont {Levitov}},\ and\ \bibinfo {author} {\bibfnamefont {T.~P.}\ \bibnamefont {Orlando}},\ }\bibfield  {title} {\bibinfo {title} {Microwave-induced cooling of a superconducting qubit},\ }\href {http://dx.doi.org/10.1126/science.1134008} {\bibfield  {journal} {\bibinfo  {journal} {Science}\ }\textbf {\bibinfo {volume} {314}},\ \bibinfo {pages} {1589} (\bibinfo {year} {2006})}\BibitemShut {NoStop}%
\bibitem [{\citenamefont {Geerlings}\ \emph {et~al.}(2013)\citenamefont {Geerlings}, \citenamefont {Leghtas}, \citenamefont {Pop}, \citenamefont {Shankar}, \citenamefont {Frunzio}, \citenamefont {Schoelkopf}, \citenamefont {Mirrahimi},\ and\ \citenamefont {Devoret}}]{Geerlings2013-kl}%
  \BibitemOpen
  \bibfield  {author} {\bibinfo {author} {\bibfnamefont {K.}~\bibnamefont {Geerlings}}, \bibinfo {author} {\bibfnamefont {Z.}~\bibnamefont {Leghtas}}, \bibinfo {author} {\bibfnamefont {I.~M.}\ \bibnamefont {Pop}}, \bibinfo {author} {\bibfnamefont {S.}~\bibnamefont {Shankar}}, \bibinfo {author} {\bibfnamefont {L.}~\bibnamefont {Frunzio}}, \bibinfo {author} {\bibfnamefont {R.~J.}\ \bibnamefont {Schoelkopf}}, \bibinfo {author} {\bibfnamefont {M.}~\bibnamefont {Mirrahimi}},\ and\ \bibinfo {author} {\bibfnamefont {M.~H.}\ \bibnamefont {Devoret}},\ }\bibfield  {title} {\bibinfo {title} {Demonstrating a driven reset protocol for a superconducting qubit},\ }\href {http://dx.doi.org/10.1103/PhysRevLett.110.120501} {\bibfield  {journal} {\bibinfo  {journal} {Phys. Rev. Lett.}\ }\textbf {\bibinfo {volume} {110}},\ \bibinfo {pages} {120501} (\bibinfo {year} {2013})}\BibitemShut {NoStop}%
\bibitem [{\citenamefont {Magnard}\ \emph {et~al.}(2018)\citenamefont {Magnard}, \citenamefont {Kurpiers}, \citenamefont {Royer}, \citenamefont {Walter}, \citenamefont {Besse}, \citenamefont {Gasparinetti}, \citenamefont {Pechal}, \citenamefont {Heinsoo}, \citenamefont {Storz}, \citenamefont {Blais},\ and\ \citenamefont {Wallraff}}]{Magnard2018-qm}%
  \BibitemOpen
  \bibfield  {author} {\bibinfo {author} {\bibfnamefont {P.}~\bibnamefont {Magnard}}, \bibinfo {author} {\bibfnamefont {P.}~\bibnamefont {Kurpiers}}, \bibinfo {author} {\bibfnamefont {B.}~\bibnamefont {Royer}}, \bibinfo {author} {\bibfnamefont {T.}~\bibnamefont {Walter}}, \bibinfo {author} {\bibfnamefont {J.-C.}\ \bibnamefont {Besse}}, \bibinfo {author} {\bibfnamefont {S.}~\bibnamefont {Gasparinetti}}, \bibinfo {author} {\bibfnamefont {M.}~\bibnamefont {Pechal}}, \bibinfo {author} {\bibfnamefont {J.}~\bibnamefont {Heinsoo}}, \bibinfo {author} {\bibfnamefont {S.}~\bibnamefont {Storz}}, \bibinfo {author} {\bibfnamefont {A.}~\bibnamefont {Blais}},\ and\ \bibinfo {author} {\bibfnamefont {A.}~\bibnamefont {Wallraff}},\ }\bibfield  {title} {\bibinfo {title} {Fast and unconditional all-microwave reset of a superconducting qubit},\ }\href {http://dx.doi.org/10.1103/PhysRevLett.121.060502} {\bibfield  {journal} {\bibinfo  {journal} {Phys. Rev. Lett.}\ }\textbf {\bibinfo {volume} {121}},\ \bibinfo {pages} {060502}
  (\bibinfo {year} {2018})}\BibitemShut {NoStop}%
\bibitem [{\citenamefont {Zhou}\ \emph {et~al.}(2021)\citenamefont {Zhou}, \citenamefont {Zhang}, \citenamefont {Yin}, \citenamefont {Huai}, \citenamefont {Gu}, \citenamefont {Xu}, \citenamefont {Allcock}, \citenamefont {Liu}, \citenamefont {Xi}, \citenamefont {Yu}, \citenamefont {Zhang}, \citenamefont {Zhang}, \citenamefont {Li}, \citenamefont {Song}, \citenamefont {Wang}, \citenamefont {Zheng}, \citenamefont {An}, \citenamefont {Zheng},\ and\ \citenamefont {Zhang}}]{Zhou2021-dc}%
  \BibitemOpen
  \bibfield  {author} {\bibinfo {author} {\bibfnamefont {Y.}~\bibnamefont {Zhou}}, \bibinfo {author} {\bibfnamefont {Z.}~\bibnamefont {Zhang}}, \bibinfo {author} {\bibfnamefont {Z.}~\bibnamefont {Yin}}, \bibinfo {author} {\bibfnamefont {S.}~\bibnamefont {Huai}}, \bibinfo {author} {\bibfnamefont {X.}~\bibnamefont {Gu}}, \bibinfo {author} {\bibfnamefont {X.}~\bibnamefont {Xu}}, \bibinfo {author} {\bibfnamefont {J.}~\bibnamefont {Allcock}}, \bibinfo {author} {\bibfnamefont {F.}~\bibnamefont {Liu}}, \bibinfo {author} {\bibfnamefont {G.}~\bibnamefont {Xi}}, \bibinfo {author} {\bibfnamefont {Q.}~\bibnamefont {Yu}}, \bibinfo {author} {\bibfnamefont {H.}~\bibnamefont {Zhang}}, \bibinfo {author} {\bibfnamefont {M.}~\bibnamefont {Zhang}}, \bibinfo {author} {\bibfnamefont {H.}~\bibnamefont {Li}}, \bibinfo {author} {\bibfnamefont {X.}~\bibnamefont {Song}}, \bibinfo {author} {\bibfnamefont {Z.}~\bibnamefont {Wang}}, \bibinfo {author} {\bibfnamefont {D.}~\bibnamefont {Zheng}}, \bibinfo {author} {\bibfnamefont
  {S.}~\bibnamefont {An}}, \bibinfo {author} {\bibfnamefont {Y.}~\bibnamefont {Zheng}},\ and\ \bibinfo {author} {\bibfnamefont {S.}~\bibnamefont {Zhang}},\ }\bibfield  {title} {\bibinfo {title} {Rapid and unconditional parametric reset protocol for tunable superconducting qubits},\ }\href {http://dx.doi.org/10.1038/s41467-021-26205-y} {\bibfield  {journal} {\bibinfo  {journal} {Nat. Commun.}\ }\textbf {\bibinfo {volume} {12}},\ \bibinfo {pages} {5924} (\bibinfo {year} {2021})}\BibitemShut {NoStop}%
\bibitem [{\citenamefont {Cao}\ \emph {et~al.}(2025)\citenamefont {Cao}, \citenamefont {Mucci}, \citenamefont {Liu}, \citenamefont {Pekker},\ and\ \citenamefont {Hatridge}}]{Cao2025-la}%
  \BibitemOpen
  \bibfield  {author} {\bibinfo {author} {\bibfnamefont {X.}~\bibnamefont {Cao}}, \bibinfo {author} {\bibfnamefont {M.}~\bibnamefont {Mucci}}, \bibinfo {author} {\bibfnamefont {G.}~\bibnamefont {Liu}}, \bibinfo {author} {\bibfnamefont {D.}~\bibnamefont {Pekker}},\ and\ \bibinfo {author} {\bibfnamefont {M.}~\bibnamefont {Hatridge}},\ }\bibfield  {title} {\bibinfo {title} {Engineering a multilevel bath for transmons with three-wave mixing and parametric drives},\ }\href {http://dx.doi.org/10.1103/mv4g-z64p} {\bibfield  {journal} {\bibinfo  {journal} {Phys. Rev. Lett.}\ }\textbf {\bibinfo {volume} {135}},\ \bibinfo {pages} {230403} (\bibinfo {year} {2025})}\BibitemShut {NoStop}%
\bibitem [{\citenamefont {Ma}\ \emph {et~al.}(2019)\citenamefont {Ma}, \citenamefont {Saxberg}, \citenamefont {Owens}, \citenamefont {Leung}, \citenamefont {Lu}, \citenamefont {Simon},\ and\ \citenamefont {Schuster}}]{Ma2019-ee}%
  \BibitemOpen
  \bibfield  {author} {\bibinfo {author} {\bibfnamefont {R.}~\bibnamefont {Ma}}, \bibinfo {author} {\bibfnamefont {B.}~\bibnamefont {Saxberg}}, \bibinfo {author} {\bibfnamefont {C.}~\bibnamefont {Owens}}, \bibinfo {author} {\bibfnamefont {N.}~\bibnamefont {Leung}}, \bibinfo {author} {\bibfnamefont {Y.}~\bibnamefont {Lu}}, \bibinfo {author} {\bibfnamefont {J.}~\bibnamefont {Simon}},\ and\ \bibinfo {author} {\bibfnamefont {D.~I.}\ \bibnamefont {Schuster}},\ }\bibfield  {title} {\bibinfo {title} {A dissipatively stabilized mott insulator of photons},\ }\href {https://www.nature.com/articles/s41586-019-0897-9} {\bibfield  {journal} {\bibinfo  {journal} {Nature}\ }\textbf {\bibinfo {volume} {566}},\ \bibinfo {pages} {51} (\bibinfo {year} {2019})}\BibitemShut {NoStop}%
\bibitem [{\citenamefont {Murch}\ \emph {et~al.}(2012)\citenamefont {Murch}, \citenamefont {Vool}, \citenamefont {Zhou}, \citenamefont {Weber}, \citenamefont {Girvin},\ and\ \citenamefont {Siddiqi}}]{Murch2012-jy}%
  \BibitemOpen
  \bibfield  {author} {\bibinfo {author} {\bibfnamefont {K.~W.}\ \bibnamefont {Murch}}, \bibinfo {author} {\bibfnamefont {U.}~\bibnamefont {Vool}}, \bibinfo {author} {\bibfnamefont {D.}~\bibnamefont {Zhou}}, \bibinfo {author} {\bibfnamefont {S.~J.}\ \bibnamefont {Weber}}, \bibinfo {author} {\bibfnamefont {S.~M.}\ \bibnamefont {Girvin}},\ and\ \bibinfo {author} {\bibfnamefont {I.}~\bibnamefont {Siddiqi}},\ }\bibfield  {title} {\bibinfo {title} {Cavity-assisted quantum bath engineering},\ }\href {http://dx.doi.org/10.1103/PhysRevLett.109.183602} {\bibfield  {journal} {\bibinfo  {journal} {Phys. Rev. Lett.}\ }\textbf {\bibinfo {volume} {109}},\ \bibinfo {pages} {183602} (\bibinfo {year} {2012})}\BibitemShut {NoStop}%
\bibitem [{\citenamefont {Lu}\ \emph {et~al.}(2017)\citenamefont {Lu}, \citenamefont {Chakram}, \citenamefont {Leung}, \citenamefont {Earnest}, \citenamefont {Naik}, \citenamefont {Huang}, \citenamefont {Groszkowski}, \citenamefont {Kapit}, \citenamefont {Koch},\ and\ \citenamefont {Schuster}}]{Lu2017-uq}%
  \BibitemOpen
  \bibfield  {author} {\bibinfo {author} {\bibfnamefont {Y.}~\bibnamefont {Lu}}, \bibinfo {author} {\bibfnamefont {S.}~\bibnamefont {Chakram}}, \bibinfo {author} {\bibfnamefont {N.}~\bibnamefont {Leung}}, \bibinfo {author} {\bibfnamefont {N.}~\bibnamefont {Earnest}}, \bibinfo {author} {\bibfnamefont {R.~K.}\ \bibnamefont {Naik}}, \bibinfo {author} {\bibfnamefont {Z.}~\bibnamefont {Huang}}, \bibinfo {author} {\bibfnamefont {P.}~\bibnamefont {Groszkowski}}, \bibinfo {author} {\bibfnamefont {E.}~\bibnamefont {Kapit}}, \bibinfo {author} {\bibfnamefont {J.}~\bibnamefont {Koch}},\ and\ \bibinfo {author} {\bibfnamefont {D.~I.}\ \bibnamefont {Schuster}},\ }\bibfield  {title} {\bibinfo {title} {Universal stabilization of a parametrically coupled qubit},\ }\href {http://dx.doi.org/10.1103/PhysRevLett.119.150502} {\bibfield  {journal} {\bibinfo  {journal} {Phys. Rev. Lett.}\ }\textbf {\bibinfo {volume} {119}},\ \bibinfo {pages} {150502} (\bibinfo {year} {2017})}\BibitemShut {NoStop}%
\bibitem [{\citenamefont {Shankar}\ \emph {et~al.}(2013)\citenamefont {Shankar}, \citenamefont {Hatridge}, \citenamefont {Leghtas}, \citenamefont {Sliwa}, \citenamefont {Narla}, \citenamefont {Vool}, \citenamefont {Girvin}, \citenamefont {Frunzio}, \citenamefont {Mirrahimi},\ and\ \citenamefont {Devoret}}]{Shankar2013-ep}%
  \BibitemOpen
  \bibfield  {author} {\bibinfo {author} {\bibfnamefont {S.}~\bibnamefont {Shankar}}, \bibinfo {author} {\bibfnamefont {M.}~\bibnamefont {Hatridge}}, \bibinfo {author} {\bibfnamefont {Z.}~\bibnamefont {Leghtas}}, \bibinfo {author} {\bibfnamefont {K.~M.}\ \bibnamefont {Sliwa}}, \bibinfo {author} {\bibfnamefont {A.}~\bibnamefont {Narla}}, \bibinfo {author} {\bibfnamefont {U.}~\bibnamefont {Vool}}, \bibinfo {author} {\bibfnamefont {S.~M.}\ \bibnamefont {Girvin}}, \bibinfo {author} {\bibfnamefont {L.}~\bibnamefont {Frunzio}}, \bibinfo {author} {\bibfnamefont {M.}~\bibnamefont {Mirrahimi}},\ and\ \bibinfo {author} {\bibfnamefont {M.~H.}\ \bibnamefont {Devoret}},\ }\bibfield  {title} {\bibinfo {title} {Autonomously stabilized entanglement between two superconducting quantum bits},\ }\href {http://dx.doi.org/10.1038/nature12802} {\bibfield  {journal} {\bibinfo  {journal} {Nature}\ }\textbf {\bibinfo {volume} {504}},\ \bibinfo {pages} {419} (\bibinfo {year} {2013})}\BibitemShut {NoStop}%
\bibitem [{\citenamefont {Li}\ \emph {et~al.}(2024)\citenamefont {Li}, \citenamefont {Roy}, \citenamefont {Lu}, \citenamefont {Kapit},\ and\ \citenamefont {Schuster}}]{Li2024-kl}%
  \BibitemOpen
  \bibfield  {author} {\bibinfo {author} {\bibfnamefont {Z.}~\bibnamefont {Li}}, \bibinfo {author} {\bibfnamefont {T.}~\bibnamefont {Roy}}, \bibinfo {author} {\bibfnamefont {Y.}~\bibnamefont {Lu}}, \bibinfo {author} {\bibfnamefont {E.}~\bibnamefont {Kapit}},\ and\ \bibinfo {author} {\bibfnamefont {D.~I.}\ \bibnamefont {Schuster}},\ }\bibfield  {title} {\bibinfo {title} {Autonomous stabilization with programmable stabilized state},\ }\href {http://dx.doi.org/10.1038/s41467-024-51262-4} {\bibfield  {journal} {\bibinfo  {journal} {Nat. Commun.}\ }\textbf {\bibinfo {volume} {15}},\ \bibinfo {pages} {6978} (\bibinfo {year} {2024})}\BibitemShut {NoStop}%
\bibitem [{\citenamefont {Brown}\ \emph {et~al.}(2022)\citenamefont {Brown}, \citenamefont {Doucet}, \citenamefont {Rist\`{e}}, \citenamefont {Ribeill}, \citenamefont {Cicak}, \citenamefont {Aumentado}, \citenamefont {Simmonds}, \citenamefont {Govia}, \citenamefont {Kamal},\ and\ \citenamefont {Ranzani}}]{Brown2022-sg}%
  \BibitemOpen
  \bibfield  {author} {\bibinfo {author} {\bibfnamefont {T.}~\bibnamefont {Brown}}, \bibinfo {author} {\bibfnamefont {E.}~\bibnamefont {Doucet}}, \bibinfo {author} {\bibfnamefont {D.}~\bibnamefont {Rist\`{e}}}, \bibinfo {author} {\bibfnamefont {G.}~\bibnamefont {Ribeill}}, \bibinfo {author} {\bibfnamefont {K.}~\bibnamefont {Cicak}}, \bibinfo {author} {\bibfnamefont {J.}~\bibnamefont {Aumentado}}, \bibinfo {author} {\bibfnamefont {R.}~\bibnamefont {Simmonds}}, \bibinfo {author} {\bibfnamefont {L.}~\bibnamefont {Govia}}, \bibinfo {author} {\bibfnamefont {A.}~\bibnamefont {Kamal}},\ and\ \bibinfo {author} {\bibfnamefont {L.}~\bibnamefont {Ranzani}},\ }\bibfield  {title} {\bibinfo {title} {Trade off-free entanglement stabilization in a superconducting qutrit-qubit system},\ }\href {http://dx.doi.org/10.1038/s41467-022-31638-0} {\bibfield  {journal} {\bibinfo  {journal} {Nat. Commun.}\ }\textbf {\bibinfo {volume} {13}},\ \bibinfo {pages} {3994} (\bibinfo {year} {2022})}\BibitemShut {NoStop}%
\bibitem [{\citenamefont {Hacohen-Gourgy}\ \emph {et~al.}(2015)\citenamefont {Hacohen-Gourgy}, \citenamefont {Ramasesh}, \citenamefont {De~Grandi}, \citenamefont {Siddiqi},\ and\ \citenamefont {Girvin}}]{Hacohen-Gourgy2015-zc}%
  \BibitemOpen
  \bibfield  {author} {\bibinfo {author} {\bibfnamefont {S.}~\bibnamefont {Hacohen-Gourgy}}, \bibinfo {author} {\bibfnamefont {V.~V.}\ \bibnamefont {Ramasesh}}, \bibinfo {author} {\bibfnamefont {C.}~\bibnamefont {De~Grandi}}, \bibinfo {author} {\bibfnamefont {I.}~\bibnamefont {Siddiqi}},\ and\ \bibinfo {author} {\bibfnamefont {S.~M.}\ \bibnamefont {Girvin}},\ }\bibfield  {title} {\bibinfo {title} {Cooling and autonomous feedback in a bose-hubbard chain with attractive interactions},\ }\href {http://arxiv.org/abs/1506.05837} {\bibfield  {journal} {\bibinfo  {journal} {Phys. Rev. Lett.}\ }\textbf {\bibinfo {volume} {115}},\ \bibinfo {pages} {240501} (\bibinfo {year} {2015})}\BibitemShut {NoStop}%
\bibitem [{\citenamefont {Chen}\ \emph {et~al.}(2025)\citenamefont {Chen}, \citenamefont {Tang}, \citenamefont {Zhou}, \citenamefont {Yi}, \citenamefont {Zhang}, \citenamefont {Zhang}, \citenamefont {Guo}, \citenamefont {Liu}, \citenamefont {Chen}, \citenamefont {Yan},\ and\ \citenamefont {Yu}}]{Chen2025-hc}%
  \BibitemOpen
  \bibfield  {author} {\bibinfo {author} {\bibfnamefont {C.}~\bibnamefont {Chen}}, \bibinfo {author} {\bibfnamefont {K.}~\bibnamefont {Tang}}, \bibinfo {author} {\bibfnamefont {Y.}~\bibnamefont {Zhou}}, \bibinfo {author} {\bibfnamefont {K.}~\bibnamefont {Yi}}, \bibinfo {author} {\bibfnamefont {X.}~\bibnamefont {Zhang}}, \bibinfo {author} {\bibfnamefont {X.}~\bibnamefont {Zhang}}, \bibinfo {author} {\bibfnamefont {H.}~\bibnamefont {Guo}}, \bibinfo {author} {\bibfnamefont {S.}~\bibnamefont {Liu}}, \bibinfo {author} {\bibfnamefont {Y.}~\bibnamefont {Chen}}, \bibinfo {author} {\bibfnamefont {T.}~\bibnamefont {Yan}},\ and\ \bibinfo {author} {\bibfnamefont {D.}~\bibnamefont {Yu}},\ }\bibfield  {title} {\bibinfo {title} {Hardware-efficient stabilization of entanglement via engineered dissipation in superconducting circuits},\ }\href {https://link.aps.org/doi/10.1103/PhysRevResearch.7.L022018} {\bibfield  {journal} {\bibinfo  {journal} {Phys. Rev. Res.}\ }\textbf {\bibinfo {volume} {7}} (\bibinfo {year}
  {2025})}\BibitemShut {NoStop}%
\bibitem [{\citenamefont {Shah}\ \emph {et~al.}(2024)\citenamefont {Shah}, \citenamefont {Yang}, \citenamefont {Joshi},\ and\ \citenamefont {Mirhosseini}}]{Shah2024-rn}%
  \BibitemOpen
  \bibfield  {author} {\bibinfo {author} {\bibfnamefont {P.~S.}\ \bibnamefont {Shah}}, \bibinfo {author} {\bibfnamefont {F.}~\bibnamefont {Yang}}, \bibinfo {author} {\bibfnamefont {C.}~\bibnamefont {Joshi}},\ and\ \bibinfo {author} {\bibfnamefont {M.}~\bibnamefont {Mirhosseini}},\ }\bibfield  {title} {\bibinfo {title} {Stabilizing remote entanglement via waveguide dissipation},\ }\href {http://dx.doi.org/10.1103/PRXQuantum.5.030346} {\bibfield  {journal} {\bibinfo  {journal} {PRX quantum}\ }\textbf {\bibinfo {volume} {5}},\ \bibinfo {pages} {030346} (\bibinfo {year} {2024})}\BibitemShut {NoStop}%
\bibitem [{\citenamefont {Grimm}\ \emph {et~al.}(2020)\citenamefont {Grimm}, \citenamefont {Frattini}, \citenamefont {Puri}, \citenamefont {Mundhada}, \citenamefont {Touzard}, \citenamefont {Mirrahimi}, \citenamefont {Girvin}, \citenamefont {Shankar},\ and\ \citenamefont {Devoret}}]{Grimm2020-mj}%
  \BibitemOpen
  \bibfield  {author} {\bibinfo {author} {\bibfnamefont {A.}~\bibnamefont {Grimm}}, \bibinfo {author} {\bibfnamefont {N.~E.}\ \bibnamefont {Frattini}}, \bibinfo {author} {\bibfnamefont {S.}~\bibnamefont {Puri}}, \bibinfo {author} {\bibfnamefont {S.~O.}\ \bibnamefont {Mundhada}}, \bibinfo {author} {\bibfnamefont {S.}~\bibnamefont {Touzard}}, \bibinfo {author} {\bibfnamefont {M.}~\bibnamefont {Mirrahimi}}, \bibinfo {author} {\bibfnamefont {S.~M.}\ \bibnamefont {Girvin}}, \bibinfo {author} {\bibfnamefont {S.}~\bibnamefont {Shankar}},\ and\ \bibinfo {author} {\bibfnamefont {M.~H.}\ \bibnamefont {Devoret}},\ }\bibfield  {title} {\bibinfo {title} {Stabilization and operation of a kerr-cat qubit},\ }\href {http://dx.doi.org/10.1038/s41586-020-2587-z} {\bibfield  {journal} {\bibinfo  {journal} {Nature}\ }\textbf {\bibinfo {volume} {584}},\ \bibinfo {pages} {205} (\bibinfo {year} {2020})}\BibitemShut {NoStop}%
\bibitem [{\citenamefont {Gertler}\ \emph {et~al.}(2021)\citenamefont {Gertler}, \citenamefont {Baker}, \citenamefont {Li}, \citenamefont {Shirol}, \citenamefont {Koch},\ and\ \citenamefont {Wang}}]{Gertler2021-jd}%
  \BibitemOpen
  \bibfield  {author} {\bibinfo {author} {\bibfnamefont {J.~M.}\ \bibnamefont {Gertler}}, \bibinfo {author} {\bibfnamefont {B.}~\bibnamefont {Baker}}, \bibinfo {author} {\bibfnamefont {J.}~\bibnamefont {Li}}, \bibinfo {author} {\bibfnamefont {S.}~\bibnamefont {Shirol}}, \bibinfo {author} {\bibfnamefont {J.}~\bibnamefont {Koch}},\ and\ \bibinfo {author} {\bibfnamefont {C.}~\bibnamefont {Wang}},\ }\bibfield  {title} {\bibinfo {title} {Protecting a bosonic qubit with autonomous quantum error correction},\ }\href {http://dx.doi.org/10.1038/s41586-021-03257-0} {\bibfield  {journal} {\bibinfo  {journal} {Nature}\ }\textbf {\bibinfo {volume} {590}},\ \bibinfo {pages} {243} (\bibinfo {year} {2021})}\BibitemShut {NoStop}%
\bibitem [{\citenamefont {Thorbeck}\ \emph {et~al.}(2024)\citenamefont {Thorbeck}, \citenamefont {McDonald}, \citenamefont {Lanes}, \citenamefont {Blair}, \citenamefont {Keefe}, \citenamefont {Stabile}, \citenamefont {Royer}, \citenamefont {Govia},\ and\ \citenamefont {Blais}}]{thorbeck2024high}%
  \BibitemOpen
  \bibfield  {author} {\bibinfo {author} {\bibfnamefont {T.}~\bibnamefont {Thorbeck}}, \bibinfo {author} {\bibfnamefont {A.}~\bibnamefont {McDonald}}, \bibinfo {author} {\bibfnamefont {O.}~\bibnamefont {Lanes}}, \bibinfo {author} {\bibfnamefont {J.}~\bibnamefont {Blair}}, \bibinfo {author} {\bibfnamefont {G.}~\bibnamefont {Keefe}}, \bibinfo {author} {\bibfnamefont {A.~A.}\ \bibnamefont {Stabile}}, \bibinfo {author} {\bibfnamefont {B.}~\bibnamefont {Royer}}, \bibinfo {author} {\bibfnamefont {L.~C.}\ \bibnamefont {Govia}},\ and\ \bibinfo {author} {\bibfnamefont {A.}~\bibnamefont {Blais}},\ }\bibfield  {title} {\bibinfo {title} {High-fidelity gates in a transmon using bath engineering for passive leakage reset},\ }\href@noop {} {\bibfield  {journal} {\bibinfo  {journal} {arXiv preprint arXiv:2411.04101}\ } (\bibinfo {year} {2024})}\BibitemShut {NoStop}%
\bibitem [{\citenamefont {Ding}\ \emph {et~al.}(2025)\citenamefont {Ding}, \citenamefont {Li}, \citenamefont {Wang}, \citenamefont {Xue}, \citenamefont {Su}, \citenamefont {Wang}, \citenamefont {Sun}, \citenamefont {Li}, \citenamefont {Zhang}, \citenamefont {Gao} \emph {et~al.}}]{ding2025multipurpose}%
  \BibitemOpen
  \bibfield  {author} {\bibinfo {author} {\bibfnamefont {J.}~\bibnamefont {Ding}}, \bibinfo {author} {\bibfnamefont {Y.}~\bibnamefont {Li}}, \bibinfo {author} {\bibfnamefont {H.}~\bibnamefont {Wang}}, \bibinfo {author} {\bibfnamefont {G.}~\bibnamefont {Xue}}, \bibinfo {author} {\bibfnamefont {T.}~\bibnamefont {Su}}, \bibinfo {author} {\bibfnamefont {C.}~\bibnamefont {Wang}}, \bibinfo {author} {\bibfnamefont {W.}~\bibnamefont {Sun}}, \bibinfo {author} {\bibfnamefont {F.}~\bibnamefont {Li}}, \bibinfo {author} {\bibfnamefont {Y.}~\bibnamefont {Zhang}}, \bibinfo {author} {\bibfnamefont {Y.}~\bibnamefont {Gao}}, \emph {et~al.},\ }\bibfield  {title} {\bibinfo {title} {Multipurpose architecture for fast reset and protective readout of superconducting qubits},\ }\href@noop {} {\bibfield  {journal} {\bibinfo  {journal} {Physical Review Applied}\ }\textbf {\bibinfo {volume} {23}},\ \bibinfo {pages} {014012} (\bibinfo {year} {2025})}\BibitemShut {NoStop}%
\bibitem [{\citenamefont {Tan}\ \emph {et~al.}(2017)\citenamefont {Tan}, \citenamefont {Partanen}, \citenamefont {Lake}, \citenamefont {Govenius}, \citenamefont {Masuda},\ and\ \citenamefont {M{\"o}tt{\"o}nen}}]{tan2017quantum}%
  \BibitemOpen
  \bibfield  {author} {\bibinfo {author} {\bibfnamefont {K.~Y.}\ \bibnamefont {Tan}}, \bibinfo {author} {\bibfnamefont {M.}~\bibnamefont {Partanen}}, \bibinfo {author} {\bibfnamefont {R.~E.}\ \bibnamefont {Lake}}, \bibinfo {author} {\bibfnamefont {J.}~\bibnamefont {Govenius}}, \bibinfo {author} {\bibfnamefont {S.}~\bibnamefont {Masuda}},\ and\ \bibinfo {author} {\bibfnamefont {M.}~\bibnamefont {M{\"o}tt{\"o}nen}},\ }\bibfield  {title} {\bibinfo {title} {Quantum-circuit refrigerator},\ }\href@noop {} {\bibfield  {journal} {\bibinfo  {journal} {Nature communications}\ }\textbf {\bibinfo {volume} {8}},\ \bibinfo {pages} {15189} (\bibinfo {year} {2017})}\BibitemShut {NoStop}%
\bibitem [{\citenamefont {Kapit}\ \emph {et~al.}(2014)\citenamefont {Kapit}, \citenamefont {Hafezi},\ and\ \citenamefont {Simon}}]{Kapit2014-zn}%
  \BibitemOpen
  \bibfield  {author} {\bibinfo {author} {\bibfnamefont {E.}~\bibnamefont {Kapit}}, \bibinfo {author} {\bibfnamefont {M.}~\bibnamefont {Hafezi}},\ and\ \bibinfo {author} {\bibfnamefont {S.~H.}\ \bibnamefont {Simon}},\ }\bibfield  {title} {\bibinfo {title} {Induced self-stabilization in fractional quantum hall states of light},\ }\href@noop {} {\bibfield  {journal} {\bibinfo  {journal} {Physical Review X}\ }\textbf {\bibinfo {volume} {4}},\ \bibinfo {pages} {31039} (\bibinfo {year} {2014})}\BibitemShut {NoStop}%
\bibitem [{\citenamefont {Hafezi}\ \emph {et~al.}(2015)\citenamefont {Hafezi}, \citenamefont {Adhikari},\ and\ \citenamefont {Taylor}}]{Hafezi2015-mw}%
  \BibitemOpen
  \bibfield  {author} {\bibinfo {author} {\bibfnamefont {M.}~\bibnamefont {Hafezi}}, \bibinfo {author} {\bibfnamefont {P.}~\bibnamefont {Adhikari}},\ and\ \bibinfo {author} {\bibfnamefont {J.~M.}\ \bibnamefont {Taylor}},\ }\bibfield  {title} {\bibinfo {title} {Chemical potential for light by parametric coupling},\ }\href {https://link.aps.org/doi/10.1103/PhysRevB.92.174305} {\bibfield  {journal} {\bibinfo  {journal} {Phys. Rev. B: Condens. Matter Mater. Phys.}\ }\textbf {\bibinfo {volume} {92}},\ \bibinfo {pages} {174305} (\bibinfo {year} {2015})}\BibitemShut {NoStop}%
\bibitem [{\citenamefont {Lebreuilly}\ \emph {et~al.}(2017)\citenamefont {Lebreuilly}, \citenamefont {Biella}, \citenamefont {Storme}, \citenamefont {Rossini}, \citenamefont {Fazio}, \citenamefont {Ciuti},\ and\ \citenamefont {Carusotto}}]{Lebreuilly2017-hu}%
  \BibitemOpen
  \bibfield  {author} {\bibinfo {author} {\bibfnamefont {J.}~\bibnamefont {Lebreuilly}}, \bibinfo {author} {\bibfnamefont {A.}~\bibnamefont {Biella}}, \bibinfo {author} {\bibfnamefont {F.}~\bibnamefont {Storme}}, \bibinfo {author} {\bibfnamefont {D.}~\bibnamefont {Rossini}}, \bibinfo {author} {\bibfnamefont {R.}~\bibnamefont {Fazio}}, \bibinfo {author} {\bibfnamefont {C.}~\bibnamefont {Ciuti}},\ and\ \bibinfo {author} {\bibfnamefont {I.}~\bibnamefont {Carusotto}},\ }\bibfield  {title} {\bibinfo {title} {Stabilizing strongly correlated photon fluids with non-markovian reservoirs},\ }\href {https://link.aps.org/doi/10.1103/PhysRevA.96.033828} {\bibfield  {journal} {\bibinfo  {journal} {Phys. Rev. A}\ }\textbf {\bibinfo {volume} {96}},\ \bibinfo {pages} {033828} (\bibinfo {year} {2017})}\BibitemShut {NoStop}%
\bibitem [{\citenamefont {Du}\ \emph {et~al.}(2024)\citenamefont {Du}, \citenamefont {Suresh}, \citenamefont {L\'{o}pez}, \citenamefont {Cadiente},\ and\ \citenamefont {Ma}}]{Du2024-xp}%
  \BibitemOpen
  \bibfield  {author} {\bibinfo {author} {\bibfnamefont {B.}~\bibnamefont {Du}}, \bibinfo {author} {\bibfnamefont {R.}~\bibnamefont {Suresh}}, \bibinfo {author} {\bibfnamefont {S.}~\bibnamefont {L\'{o}pez}}, \bibinfo {author} {\bibfnamefont {J.}~\bibnamefont {Cadiente}},\ and\ \bibinfo {author} {\bibfnamefont {R.}~\bibnamefont {Ma}},\ }\bibfield  {title} {\bibinfo {title} {Probing site-resolved current in strongly interacting superconducting circuit lattices},\ }\href {http://link.aps.org/pdf/10.1103/PhysRevLett.133.060601} {\bibfield  {journal} {\bibinfo  {journal} {Phys. Rev. Lett.}\ }\textbf {\bibinfo {volume} {133}},\ \bibinfo {pages} {060601} (\bibinfo {year} {2024})}\BibitemShut {NoStop}%
\bibitem [{\citenamefont {Du}\ \emph {et~al.}(2025)\citenamefont {Du}, \citenamefont {Guo}, \citenamefont {L\'{o}pez},\ and\ \citenamefont {Ma}}]{Du2025-iw}%
  \BibitemOpen
  \bibfield  {author} {\bibinfo {author} {\bibfnamefont {B.}~\bibnamefont {Du}}, \bibinfo {author} {\bibfnamefont {Q.}~\bibnamefont {Guo}}, \bibinfo {author} {\bibfnamefont {S.}~\bibnamefont {L\'{o}pez}},\ and\ \bibinfo {author} {\bibfnamefont {R.}~\bibnamefont {Ma}},\ }\bibfield  {title} {\bibinfo {title} {Tunneling spectroscopy in superconducting circuit lattices},\ }\href {http://dx.doi.org/10.1103/PhysRevResearch.7.L022038} {\bibfield  {journal} {\bibinfo  {journal} {Phys. Rev. Res.}\ }\textbf {\bibinfo {volume} {7}},\ \bibinfo {pages} {L022038} (\bibinfo {year} {2025})}\BibitemShut {NoStop}%
\bibitem [{\citenamefont {Cazalilla}\ \emph {et~al.}(2011)\citenamefont {Cazalilla}, \citenamefont {Citro}, \citenamefont {Giamarchi}, \citenamefont {Orignac},\ and\ \citenamefont {Rigol}}]{cazalilla2011one}%
  \BibitemOpen
  \bibfield  {author} {\bibinfo {author} {\bibfnamefont {M.~A.}\ \bibnamefont {Cazalilla}}, \bibinfo {author} {\bibfnamefont {R.}~\bibnamefont {Citro}}, \bibinfo {author} {\bibfnamefont {T.}~\bibnamefont {Giamarchi}}, \bibinfo {author} {\bibfnamefont {E.}~\bibnamefont {Orignac}},\ and\ \bibinfo {author} {\bibfnamefont {M.}~\bibnamefont {Rigol}},\ }\bibfield  {title} {\bibinfo {title} {One dimensional bosons: From condensed matter systems to ultracold gases},\ }\href@noop {} {\bibfield  {journal} {\bibinfo  {journal} {Reviews of Modern Physics}\ }\textbf {\bibinfo {volume} {83}},\ \bibinfo {pages} {1405} (\bibinfo {year} {2011})}\BibitemShut {NoStop}%
\bibitem [{\citenamefont {Yan}\ \emph {et~al.}(2019)\citenamefont {Yan}, \citenamefont {Zhang}, \citenamefont {Gong}, \citenamefont {Wu}, \citenamefont {Zheng}, \citenamefont {Li}, \citenamefont {Wang}, \citenamefont {Liang}, \citenamefont {Lin}, \citenamefont {Xu} \emph {et~al.}}]{yan2019strongly}%
  \BibitemOpen
  \bibfield  {author} {\bibinfo {author} {\bibfnamefont {Z.}~\bibnamefont {Yan}}, \bibinfo {author} {\bibfnamefont {Y.-R.}\ \bibnamefont {Zhang}}, \bibinfo {author} {\bibfnamefont {M.}~\bibnamefont {Gong}}, \bibinfo {author} {\bibfnamefont {Y.}~\bibnamefont {Wu}}, \bibinfo {author} {\bibfnamefont {Y.}~\bibnamefont {Zheng}}, \bibinfo {author} {\bibfnamefont {S.}~\bibnamefont {Li}}, \bibinfo {author} {\bibfnamefont {C.}~\bibnamefont {Wang}}, \bibinfo {author} {\bibfnamefont {F.}~\bibnamefont {Liang}}, \bibinfo {author} {\bibfnamefont {J.}~\bibnamefont {Lin}}, \bibinfo {author} {\bibfnamefont {Y.}~\bibnamefont {Xu}}, \emph {et~al.},\ }\bibfield  {title} {\bibinfo {title} {Strongly correlated quantum walks with a 12-qubit superconducting processor},\ }\href@noop {} {\bibfield  {journal} {\bibinfo  {journal} {Science}\ }\textbf {\bibinfo {volume} {364}},\ \bibinfo {pages} {753} (\bibinfo {year} {2019})}\BibitemShut {NoStop}%
\bibitem [{\citenamefont {Karamlou}\ \emph {et~al.}(2024)\citenamefont {Karamlou}, \citenamefont {Rosen}, \citenamefont {Muschinske}, \citenamefont {Barrett}, \citenamefont {Di~Paolo}, \citenamefont {Ding}, \citenamefont {Harrington}, \citenamefont {Hays}, \citenamefont {Das}, \citenamefont {Kim}, \citenamefont {Niedzielski}, \citenamefont {Schuldt}, \citenamefont {Serniak}, \citenamefont {Schwartz}, \citenamefont {Yoder}, \citenamefont {Gustavsson}, \citenamefont {Yanay}, \citenamefont {Grover},\ and\ \citenamefont {Oliver}}]{Karamlou2024-gt}%
  \BibitemOpen
  \bibfield  {author} {\bibinfo {author} {\bibfnamefont {A.~H.}\ \bibnamefont {Karamlou}}, \bibinfo {author} {\bibfnamefont {I.~T.}\ \bibnamefont {Rosen}}, \bibinfo {author} {\bibfnamefont {S.~E.}\ \bibnamefont {Muschinske}}, \bibinfo {author} {\bibfnamefont {C.~N.}\ \bibnamefont {Barrett}}, \bibinfo {author} {\bibfnamefont {A.}~\bibnamefont {Di~Paolo}}, \bibinfo {author} {\bibfnamefont {L.}~\bibnamefont {Ding}}, \bibinfo {author} {\bibfnamefont {P.~M.}\ \bibnamefont {Harrington}}, \bibinfo {author} {\bibfnamefont {M.}~\bibnamefont {Hays}}, \bibinfo {author} {\bibfnamefont {R.}~\bibnamefont {Das}}, \bibinfo {author} {\bibfnamefont {D.~K.}\ \bibnamefont {Kim}}, \bibinfo {author} {\bibfnamefont {B.~M.}\ \bibnamefont {Niedzielski}}, \bibinfo {author} {\bibfnamefont {M.}~\bibnamefont {Schuldt}}, \bibinfo {author} {\bibfnamefont {K.}~\bibnamefont {Serniak}}, \bibinfo {author} {\bibfnamefont {M.~E.}\ \bibnamefont {Schwartz}}, \bibinfo {author} {\bibfnamefont {J.~L.}\ \bibnamefont {Yoder}}, \bibinfo {author}
  {\bibfnamefont {S.}~\bibnamefont {Gustavsson}}, \bibinfo {author} {\bibfnamefont {Y.}~\bibnamefont {Yanay}}, \bibinfo {author} {\bibfnamefont {J.~A.}\ \bibnamefont {Grover}},\ and\ \bibinfo {author} {\bibfnamefont {W.~D.}\ \bibnamefont {Oliver}},\ }\bibfield  {title} {\bibinfo {title} {Probing entanglement in a {2D} hard-core bose-hubbard lattice},\ }\href {http://dx.doi.org/10.1038/s41586-024-07325-z} {\bibfield  {journal} {\bibinfo  {journal} {Nature}\ }\textbf {\bibinfo {volume} {629}},\ \bibinfo {pages} {561} (\bibinfo {year} {2024})}\BibitemShut {NoStop}%
\bibitem [{\citenamefont {Huang}\ \emph {et~al.}(2020)\citenamefont {Huang}, \citenamefont {Kueng},\ and\ \citenamefont {Preskill}}]{Huang2020-cd}%
  \BibitemOpen
  \bibfield  {author} {\bibinfo {author} {\bibfnamefont {H.-Y.}\ \bibnamefont {Huang}}, \bibinfo {author} {\bibfnamefont {R.}~\bibnamefont {Kueng}},\ and\ \bibinfo {author} {\bibfnamefont {J.}~\bibnamefont {Preskill}},\ }\bibfield  {title} {\bibinfo {title} {Predicting many properties of a quantum system from very few measurements},\ }\href {https://www.nature.com/articles/s41567-020-0932-7} {\bibfield  {journal} {\bibinfo  {journal} {Nat. Phys.}\ }\textbf {\bibinfo {volume} {16}},\ \bibinfo {pages} {1050} (\bibinfo {year} {2020})}\BibitemShut {NoStop}%
\bibitem [{\citenamefont {Chen}\ \emph {et~al.}(2021)\citenamefont {Chen}, \citenamefont {Yu}, \citenamefont {Zeng},\ and\ \citenamefont {Flammia}}]{Chen2021-cg}%
  \BibitemOpen
  \bibfield  {author} {\bibinfo {author} {\bibfnamefont {S.}~\bibnamefont {Chen}}, \bibinfo {author} {\bibfnamefont {W.}~\bibnamefont {Yu}}, \bibinfo {author} {\bibfnamefont {P.}~\bibnamefont {Zeng}},\ and\ \bibinfo {author} {\bibfnamefont {S.~T.}\ \bibnamefont {Flammia}},\ }\bibfield  {title} {\bibinfo {title} {Robust shadow estimation},\ }\href {http://dx.doi.org/10.1103/prxquantum.2.030348} {\bibfield  {journal} {\bibinfo  {journal} {PRX Quantum}\ }\textbf {\bibinfo {volume} {2}},\ \bibinfo {pages} {030348} (\bibinfo {year} {2021})}\BibitemShut {NoStop}%
\bibitem [{\citenamefont {Vermersch}\ \emph {et~al.}(2024)\citenamefont {Vermersch}, \citenamefont {Rath}, \citenamefont {Sundar}, \citenamefont {Branciard}, \citenamefont {Preskill},\ and\ \citenamefont {Elben}}]{Vermersch2024-tv}%
  \BibitemOpen
  \bibfield  {author} {\bibinfo {author} {\bibfnamefont {B.}~\bibnamefont {Vermersch}}, \bibinfo {author} {\bibfnamefont {A.}~\bibnamefont {Rath}}, \bibinfo {author} {\bibfnamefont {B.}~\bibnamefont {Sundar}}, \bibinfo {author} {\bibfnamefont {C.}~\bibnamefont {Branciard}}, \bibinfo {author} {\bibfnamefont {J.}~\bibnamefont {Preskill}},\ and\ \bibinfo {author} {\bibfnamefont {A.}~\bibnamefont {Elben}},\ }\bibfield  {title} {\bibinfo {title} {Enhanced estimation of quantum properties with common randomized measurements},\ }\href {http://dx.doi.org/10.1103/PRXQuantum.5.010352} {\bibfield  {journal} {\bibinfo  {journal} {PRX quantum}\ }\textbf {\bibinfo {volume} {5}},\ \bibinfo {pages} {010352} (\bibinfo {year} {2024})}\BibitemShut {NoStop}%
\bibitem [{\citenamefont {Vitale}\ \emph {et~al.}(2024)\citenamefont {Vitale}, \citenamefont {Rath}, \citenamefont {Jurcevic}, \citenamefont {Elben}, \citenamefont {Branciard},\ and\ \citenamefont {Vermersch}}]{Vitale2024-oa}%
  \BibitemOpen
  \bibfield  {author} {\bibinfo {author} {\bibfnamefont {V.}~\bibnamefont {Vitale}}, \bibinfo {author} {\bibfnamefont {A.}~\bibnamefont {Rath}}, \bibinfo {author} {\bibfnamefont {P.}~\bibnamefont {Jurcevic}}, \bibinfo {author} {\bibfnamefont {A.}~\bibnamefont {Elben}}, \bibinfo {author} {\bibfnamefont {C.}~\bibnamefont {Branciard}},\ and\ \bibinfo {author} {\bibfnamefont {B.}~\bibnamefont {Vermersch}},\ }\bibfield  {title} {\bibinfo {title} {Robust estimation of the quantum fisher information on a quantum processor},\ }\href {http://dx.doi.org/10.1103/PRXQuantum.5.030338} {\bibfield  {journal} {\bibinfo  {journal} {PRX quantum}\ }\textbf {\bibinfo {volume} {5}},\ \bibinfo {pages} {030338} (\bibinfo {year} {2024})}\BibitemShut {NoStop}%
\bibitem [{\citenamefont {Wootters}(1998)}]{Wootters1998-rd}%
  \BibitemOpen
  \bibfield  {author} {\bibinfo {author} {\bibfnamefont {W.~K.}\ \bibnamefont {Wootters}},\ }\bibfield  {title} {\bibinfo {title} {Entanglement of formation of an arbitrary state of two qubits},\ }\href {http://dx.doi.org/10.1103/PhysRevLett.80.2245} {\bibfield  {journal} {\bibinfo  {journal} {Phys. Rev. Lett.}\ }\textbf {\bibinfo {volume} {80}},\ \bibinfo {pages} {2245} (\bibinfo {year} {1998})}\BibitemShut {NoStop}%
\bibitem [{\citenamefont {Levine}\ \emph {et~al.}(2024)\citenamefont {Levine}, \citenamefont {Haim}, \citenamefont {Hung}, \citenamefont {Alidoust}, \citenamefont {Kalaee}, \citenamefont {DeLorenzo}, \citenamefont {Wollack}, \citenamefont {Arrangoiz-Arriola}, \citenamefont {Khalajhedayati}, \citenamefont {Sanil} \emph {et~al.}}]{Levine2024-yd_truncate}%
  \BibitemOpen
  \bibfield  {author} {\bibinfo {author} {\bibfnamefont {H.}~\bibnamefont {Levine}}, \bibinfo {author} {\bibfnamefont {A.}~\bibnamefont {Haim}}, \bibinfo {author} {\bibfnamefont {J.~S.~C.}\ \bibnamefont {Hung}}, \bibinfo {author} {\bibfnamefont {N.}~\bibnamefont {Alidoust}}, \bibinfo {author} {\bibfnamefont {M.}~\bibnamefont {Kalaee}}, \bibinfo {author} {\bibfnamefont {L.}~\bibnamefont {DeLorenzo}}, \bibinfo {author} {\bibfnamefont {E.~A.}\ \bibnamefont {Wollack}}, \bibinfo {author} {\bibfnamefont {P.}~\bibnamefont {Arrangoiz-Arriola}}, \bibinfo {author} {\bibfnamefont {A.}~\bibnamefont {Khalajhedayati}}, \bibinfo {author} {\bibfnamefont {R.}~\bibnamefont {Sanil}}, \emph {et~al.},\ }\bibfield  {title} {\bibinfo {title} {Demonstrating a long-coherence dual-rail erasure qubit using tunable transmons},\ }\href {https://link.aps.org/doi/10.1103/PhysRevX.14.011051} {\bibfield  {journal} {\bibinfo  {journal} {Phys. Rev. X}\ }\textbf {\bibinfo {volume} {14}},\ \bibinfo {pages} {011051} (\bibinfo {year}
  {2024})}\BibitemShut {NoStop}%
\bibitem [{\citenamefont {Yeh}\ \emph {et~al.}(2017)\citenamefont {Yeh}, \citenamefont {LeFebvre}, \citenamefont {Premaratne}, \citenamefont {Wellstood},\ and\ \citenamefont {Palmer}}]{yeh2017microwave}%
  \BibitemOpen
  \bibfield  {author} {\bibinfo {author} {\bibfnamefont {J.-H.}\ \bibnamefont {Yeh}}, \bibinfo {author} {\bibfnamefont {J.}~\bibnamefont {LeFebvre}}, \bibinfo {author} {\bibfnamefont {S.}~\bibnamefont {Premaratne}}, \bibinfo {author} {\bibfnamefont {F.}~\bibnamefont {Wellstood}},\ and\ \bibinfo {author} {\bibfnamefont {B.}~\bibnamefont {Palmer}},\ }\bibfield  {title} {\bibinfo {title} {Microwave attenuators for use with quantum devices below 100 mk},\ }\href@noop {} {\bibfield  {journal} {\bibinfo  {journal} {Journal of Applied Physics}\ }\textbf {\bibinfo {volume} {121}} (\bibinfo {year} {2017})}\BibitemShut {NoStop}%
\bibitem [{\citenamefont {{Z. Wang, S. Shankar, Z. K. Minev, P. Campagne-Ibarcq, A. Narla, and M. H. Devoret}}(2019)}]{PhysRevApplied.11.014031}%
  \BibitemOpen
  \bibfield  {author} {\bibinfo {author} {\bibnamefont {{Z. Wang, S. Shankar, Z. K. Minev, P. Campagne-Ibarcq, A. Narla, and M. H. Devoret}}},\ }\bibfield  {title} {\bibinfo {title} {Cavity attenuators for superconducting qubits},\ }\href {https://doi.org/10.1103/PhysRevApplied.11.014031} {\bibfield  {journal} {\bibinfo  {journal} {Phys. Rev. Appl.}\ }\textbf {\bibinfo {volume} {11}},\ \bibinfo {pages} {014031} (\bibinfo {year} {2019})}\BibitemShut {NoStop}%
\bibitem [{\citenamefont {Krinner}\ \emph {et~al.}(2019)\citenamefont {Krinner}, \citenamefont {Storz}, \citenamefont {Kurpiers}, \citenamefont {Magnard}, \citenamefont {Heinsoo}, \citenamefont {Keller}, \citenamefont {Luetolf}, \citenamefont {Eichler},\ and\ \citenamefont {Wallraff}}]{krinner2019engineering}%
  \BibitemOpen
  \bibfield  {author} {\bibinfo {author} {\bibfnamefont {S.}~\bibnamefont {Krinner}}, \bibinfo {author} {\bibfnamefont {S.}~\bibnamefont {Storz}}, \bibinfo {author} {\bibfnamefont {P.}~\bibnamefont {Kurpiers}}, \bibinfo {author} {\bibfnamefont {P.}~\bibnamefont {Magnard}}, \bibinfo {author} {\bibfnamefont {J.}~\bibnamefont {Heinsoo}}, \bibinfo {author} {\bibfnamefont {R.}~\bibnamefont {Keller}}, \bibinfo {author} {\bibfnamefont {J.}~\bibnamefont {Luetolf}}, \bibinfo {author} {\bibfnamefont {C.}~\bibnamefont {Eichler}},\ and\ \bibinfo {author} {\bibfnamefont {A.}~\bibnamefont {Wallraff}},\ }\bibfield  {title} {\bibinfo {title} {Engineering cryogenic setups for 100-qubit scale superconducting circuit systems},\ }\href@noop {} {\bibfield  {journal} {\bibinfo  {journal} {EPJ Quantum technology}\ }\textbf {\bibinfo {volume} {6}},\ \bibinfo {pages} {2} (\bibinfo {year} {2019})}\BibitemShut {NoStop}%
\bibitem [{\citenamefont {Place}\ \emph {et~al.}(2021)\citenamefont {Place}, \citenamefont {Rodgers}, \citenamefont {Mundada}, \citenamefont {Smitham}, \citenamefont {Fitzpatrick}, \citenamefont {Leng}, \citenamefont {Premkumar}, \citenamefont {Bryon}, \citenamefont {Vrajitoarea}, \citenamefont {Sussman} \emph {et~al.}}]{place2021new}%
  \BibitemOpen
  \bibfield  {author} {\bibinfo {author} {\bibfnamefont {A.~P.}\ \bibnamefont {Place}}, \bibinfo {author} {\bibfnamefont {L.~V.}\ \bibnamefont {Rodgers}}, \bibinfo {author} {\bibfnamefont {P.}~\bibnamefont {Mundada}}, \bibinfo {author} {\bibfnamefont {B.~M.}\ \bibnamefont {Smitham}}, \bibinfo {author} {\bibfnamefont {M.}~\bibnamefont {Fitzpatrick}}, \bibinfo {author} {\bibfnamefont {Z.}~\bibnamefont {Leng}}, \bibinfo {author} {\bibfnamefont {A.}~\bibnamefont {Premkumar}}, \bibinfo {author} {\bibfnamefont {J.}~\bibnamefont {Bryon}}, \bibinfo {author} {\bibfnamefont {A.}~\bibnamefont {Vrajitoarea}}, \bibinfo {author} {\bibfnamefont {S.}~\bibnamefont {Sussman}}, \emph {et~al.},\ }\bibfield  {title} {\bibinfo {title} {New material platform for superconducting transmon qubits with coherence times exceeding 0.3 milliseconds},\ }\href@noop {} {\bibfield  {journal} {\bibinfo  {journal} {Nature communications}\ }\textbf {\bibinfo {volume} {12}},\ \bibinfo {pages} {1779} (\bibinfo {year} {2021})}\BibitemShut {NoStop}%
\bibitem [{\citenamefont {Wang}\ \emph {et~al.}(2022)\citenamefont {Wang}, \citenamefont {Li}, \citenamefont {Xu}, \citenamefont {Li}, \citenamefont {Wang}, \citenamefont {Yang}, \citenamefont {Mi}, \citenamefont {Liang}, \citenamefont {Su}, \citenamefont {Yang} \emph {et~al.}}]{wang2022towards}%
  \BibitemOpen
  \bibfield  {author} {\bibinfo {author} {\bibfnamefont {C.}~\bibnamefont {Wang}}, \bibinfo {author} {\bibfnamefont {X.}~\bibnamefont {Li}}, \bibinfo {author} {\bibfnamefont {H.}~\bibnamefont {Xu}}, \bibinfo {author} {\bibfnamefont {Z.}~\bibnamefont {Li}}, \bibinfo {author} {\bibfnamefont {J.}~\bibnamefont {Wang}}, \bibinfo {author} {\bibfnamefont {Z.}~\bibnamefont {Yang}}, \bibinfo {author} {\bibfnamefont {Z.}~\bibnamefont {Mi}}, \bibinfo {author} {\bibfnamefont {X.}~\bibnamefont {Liang}}, \bibinfo {author} {\bibfnamefont {T.}~\bibnamefont {Su}}, \bibinfo {author} {\bibfnamefont {C.}~\bibnamefont {Yang}}, \emph {et~al.},\ }\bibfield  {title} {\bibinfo {title} {Towards practical quantum computers: Transmon qubit with a lifetime approaching 0.5 milliseconds},\ }\href@noop {} {\bibfield  {journal} {\bibinfo  {journal} {npj Quantum Information}\ }\textbf {\bibinfo {volume} {8}},\ \bibinfo {pages} {3} (\bibinfo {year} {2022})}\BibitemShut {NoStop}%
\bibitem [{\citenamefont {Jeffrey}\ \emph {et~al.}(2014)\citenamefont {Jeffrey}, \citenamefont {Sank}, \citenamefont {Mutus}, \citenamefont {White}, \citenamefont {Kelly}, \citenamefont {Barends}, \citenamefont {Chen}, \citenamefont {Chen}, \citenamefont {Chiaro}, \citenamefont {Dunsworth}, \citenamefont {Megrant}, \citenamefont {O'Malley}, \citenamefont {Neill}, \citenamefont {Roushan}, \citenamefont {Vainsencher}, \citenamefont {Wenner}, \citenamefont {Cleland},\ and\ \citenamefont {Martinis}}]{PhysRevLett.112.190504}%
  \BibitemOpen
  \bibfield  {author} {\bibinfo {author} {\bibfnamefont {E.}~\bibnamefont {Jeffrey}}, \bibinfo {author} {\bibfnamefont {D.}~\bibnamefont {Sank}}, \bibinfo {author} {\bibfnamefont {J.~Y.}\ \bibnamefont {Mutus}}, \bibinfo {author} {\bibfnamefont {T.~C.}\ \bibnamefont {White}}, \bibinfo {author} {\bibfnamefont {J.}~\bibnamefont {Kelly}}, \bibinfo {author} {\bibfnamefont {R.}~\bibnamefont {Barends}}, \bibinfo {author} {\bibfnamefont {Y.}~\bibnamefont {Chen}}, \bibinfo {author} {\bibfnamefont {Z.}~\bibnamefont {Chen}}, \bibinfo {author} {\bibfnamefont {B.}~\bibnamefont {Chiaro}}, \bibinfo {author} {\bibfnamefont {A.}~\bibnamefont {Dunsworth}}, \bibinfo {author} {\bibfnamefont {A.}~\bibnamefont {Megrant}}, \bibinfo {author} {\bibfnamefont {P.~J.~J.}\ \bibnamefont {O'Malley}}, \bibinfo {author} {\bibfnamefont {C.}~\bibnamefont {Neill}}, \bibinfo {author} {\bibfnamefont {P.}~\bibnamefont {Roushan}}, \bibinfo {author} {\bibfnamefont {A.}~\bibnamefont {Vainsencher}}, \bibinfo {author} {\bibfnamefont {J.}~\bibnamefont
  {Wenner}}, \bibinfo {author} {\bibfnamefont {A.~N.}\ \bibnamefont {Cleland}},\ and\ \bibinfo {author} {\bibfnamefont {J.~M.}\ \bibnamefont {Martinis}},\ }\bibfield  {title} {\bibinfo {title} {Fast accurate state measurement with superconducting qubits},\ }\href {https://doi.org/10.1103/PhysRevLett.112.190504} {\bibfield  {journal} {\bibinfo  {journal} {Phys. Rev. Lett.}\ }\textbf {\bibinfo {volume} {112}},\ \bibinfo {pages} {190504} (\bibinfo {year} {2014})}\BibitemShut {NoStop}%
\bibitem [{\citenamefont {Chu}\ \emph {et~al.}(2025)\citenamefont {Chu}, \citenamefont {Mamaev}, \citenamefont {Koppenh{\"{o}}fer}, \citenamefont {Yuan},\ and\ \citenamefont {Clerk}}]{Chu2025-vx}%
  \BibitemOpen
  \bibfield  {author} {\bibinfo {author} {\bibfnamefont {A.}~\bibnamefont {Chu}}, \bibinfo {author} {\bibfnamefont {M.}~\bibnamefont {Mamaev}}, \bibinfo {author} {\bibfnamefont {M.}~\bibnamefont {Koppenh{\"{o}}fer}}, \bibinfo {author} {\bibfnamefont {M.}~\bibnamefont {Yuan}},\ and\ \bibinfo {author} {\bibfnamefont {A.~A.}\ \bibnamefont {Clerk}},\ }\bibfield  {title} {\bibinfo {title} {Reconfigurable dissipative entanglement between many spin ensembles: from robust quantum sensing to many-body state engineering},\ }\href {http://arxiv.org/abs/2510.07616} {\bibfield  {journal} {\bibinfo  {journal} {arXiv [quant-ph]}\ } (\bibinfo {year} {2025})}\BibitemShut {NoStop}%
\bibitem [{\citenamefont {Guo}\ \emph {et~al.}(2025)\citenamefont {Guo}, \citenamefont {Hart}, \citenamefont {Chen}, \citenamefont {Friedman},\ and\ \citenamefont {Lucas}}]{Guo2025-rj}%
  \BibitemOpen
  \bibfield  {author} {\bibinfo {author} {\bibfnamefont {J.}~\bibnamefont {Guo}}, \bibinfo {author} {\bibfnamefont {O.}~\bibnamefont {Hart}}, \bibinfo {author} {\bibfnamefont {C.-F.}\ \bibnamefont {Chen}}, \bibinfo {author} {\bibfnamefont {A.~J.}\ \bibnamefont {Friedman}},\ and\ \bibinfo {author} {\bibfnamefont {A.}~\bibnamefont {Lucas}},\ }\bibfield  {title} {\bibinfo {title} {Designing open quantum systems with known steady states: Davies generators and beyond},\ }\href {http://dx.doi.org/10.22331/q-2025-01-28-1612} {\bibfield  {journal} {\bibinfo  {journal} {Quantum}\ }\textbf {\bibinfo {volume} {9}},\ \bibinfo {pages} {1612} (\bibinfo {year} {2025})}\BibitemShut {NoStop}%
\bibitem [{\citenamefont {Mi}\ \emph {et~al.}(2024)\citenamefont {Mi}, \citenamefont {Michailidis}, \citenamefont {Shabani}, \citenamefont {Miao}, \citenamefont {Klimov}, \citenamefont {Lloyd}, \citenamefont {Rosenberg}, \citenamefont {Acharya}, \citenamefont {Aleiner}, \citenamefont {Andersen} \emph {et~al.}}]{Mi2024-in_truncate}%
  \BibitemOpen
  \bibfield  {author} {\bibinfo {author} {\bibfnamefont {X.}~\bibnamefont {Mi}}, \bibinfo {author} {\bibfnamefont {A.~A.}\ \bibnamefont {Michailidis}}, \bibinfo {author} {\bibfnamefont {S.}~\bibnamefont {Shabani}}, \bibinfo {author} {\bibfnamefont {K.~C.}\ \bibnamefont {Miao}}, \bibinfo {author} {\bibfnamefont {P.~V.}\ \bibnamefont {Klimov}}, \bibinfo {author} {\bibfnamefont {J.}~\bibnamefont {Lloyd}}, \bibinfo {author} {\bibfnamefont {E.}~\bibnamefont {Rosenberg}}, \bibinfo {author} {\bibfnamefont {R.}~\bibnamefont {Acharya}}, \bibinfo {author} {\bibfnamefont {I.}~\bibnamefont {Aleiner}}, \bibinfo {author} {\bibfnamefont {T.~I.}\ \bibnamefont {Andersen}}, \emph {et~al.},\ }\bibfield  {title} {\bibinfo {title} {Stable quantum-correlated many-body states through engineered dissipation},\ }\href {http://dx.doi.org/10.1126/science.adh9932} {\bibfield  {journal} {\bibinfo  {journal} {Science}\ }\textbf {\bibinfo {volume} {383}},\ \bibinfo {pages} {1332} (\bibinfo {year} {2024})}\BibitemShut {NoStop}%
\bibitem [{\citenamefont {Zhan}\ \emph {et~al.}(2025)\citenamefont {Zhan}, \citenamefont {Ding}, \citenamefont {Huhn}, \citenamefont {Gray}, \citenamefont {Preskill}, \citenamefont {Chan},\ and\ \citenamefont {Lin}}]{Zhan2025-xx}%
  \BibitemOpen
  \bibfield  {author} {\bibinfo {author} {\bibfnamefont {Y.}~\bibnamefont {Zhan}}, \bibinfo {author} {\bibfnamefont {Z.}~\bibnamefont {Ding}}, \bibinfo {author} {\bibfnamefont {J.}~\bibnamefont {Huhn}}, \bibinfo {author} {\bibfnamefont {J.}~\bibnamefont {Gray}}, \bibinfo {author} {\bibfnamefont {J.}~\bibnamefont {Preskill}}, \bibinfo {author} {\bibfnamefont {G.~K.-L.}\ \bibnamefont {Chan}},\ and\ \bibinfo {author} {\bibfnamefont {L.}~\bibnamefont {Lin}},\ }\bibfield  {title} {\bibinfo {title} {Rapid quantum ground state preparation via dissipative dynamics},\ }\href {http://arxiv.org/abs/2503.15827} {\bibfield  {journal} {\bibinfo  {journal} {arXiv [quant-ph]}\ } (\bibinfo {year} {2025})}\BibitemShut {NoStop}%
\bibitem [{\citenamefont {Marcos}\ \emph {et~al.}(2012)\citenamefont {Marcos}, \citenamefont {Tomadin}, \citenamefont {Diehl},\ and\ \citenamefont {Rabl}}]{Marcos2012-ep}%
  \BibitemOpen
  \bibfield  {author} {\bibinfo {author} {\bibfnamefont {D.}~\bibnamefont {Marcos}}, \bibinfo {author} {\bibfnamefont {A.}~\bibnamefont {Tomadin}}, \bibinfo {author} {\bibfnamefont {S.}~\bibnamefont {Diehl}},\ and\ \bibinfo {author} {\bibfnamefont {P.}~\bibnamefont {Rabl}},\ }\bibfield  {title} {\bibinfo {title} {Photon condensation in circuit quantum electrodynamics by engineered dissipation},\ }\href {http://www.njp.org/} {\bibfield  {journal} {\bibinfo  {journal} {New J. Phys.}\ }\textbf {\bibinfo {volume} {14}},\ \bibinfo {pages} {055005} (\bibinfo {year} {2012})}\BibitemShut {NoStop}%
\bibitem [{\citenamefont {Lebreuilly}\ \emph {et~al.}(2016)\citenamefont {Lebreuilly}, \citenamefont {Wouters},\ and\ \citenamefont {Carusotto}}]{Lebreuilly2016-py}%
  \BibitemOpen
  \bibfield  {author} {\bibinfo {author} {\bibfnamefont {J.}~\bibnamefont {Lebreuilly}}, \bibinfo {author} {\bibfnamefont {M.}~\bibnamefont {Wouters}},\ and\ \bibinfo {author} {\bibfnamefont {I.}~\bibnamefont {Carusotto}},\ }\bibfield  {title} {\bibinfo {title} {Towards strongly correlated photons in arrays of dissipative nonlinear cavities under a frequency-dependent incoherent pumping},\ }\href {https://doi.org/10.1016/j.crhy.2016.07.001} {\bibfield  {journal} {\bibinfo  {journal} {C. R. Phys.}\ }\textbf {\bibinfo {volume} {17}},\ \bibinfo {pages} {836} (\bibinfo {year} {2016})}\BibitemShut {NoStop}%
\bibitem [{\citenamefont {Ma}\ \emph {et~al.}(2017)\citenamefont {Ma}, \citenamefont {Owens}, \citenamefont {Houck}, \citenamefont {Schuster},\ and\ \citenamefont {Simon}}]{Ma2017-vw}%
  \BibitemOpen
  \bibfield  {author} {\bibinfo {author} {\bibfnamefont {R.}~\bibnamefont {Ma}}, \bibinfo {author} {\bibfnamefont {C.}~\bibnamefont {Owens}}, \bibinfo {author} {\bibfnamefont {A.}~\bibnamefont {Houck}}, \bibinfo {author} {\bibfnamefont {D.~I.}\ \bibnamefont {Schuster}},\ and\ \bibinfo {author} {\bibfnamefont {J.}~\bibnamefont {Simon}},\ }\bibfield  {title} {\bibinfo {title} {Autonomous stabilizer for incompressible photon fluids and solids},\ }\href {https://link.aps.org/doi/10.1103/PhysRevA.95.043811} {\bibfield  {journal} {\bibinfo  {journal} {Phys. Rev. A}\ }\textbf {\bibinfo {volume} {95}},\ \bibinfo {pages} {043811} (\bibinfo {year} {2017})}\BibitemShut {NoStop}%
\bibitem [{\citenamefont {Kurilovich}\ \emph {et~al.}(2022)\citenamefont {Kurilovich}, \citenamefont {Kurilovich}, \citenamefont {Lebreuilly},\ and\ \citenamefont {Girvin}}]{Kurilovich2022-zj}%
  \BibitemOpen
  \bibfield  {author} {\bibinfo {author} {\bibfnamefont {P.}~\bibnamefont {Kurilovich}}, \bibinfo {author} {\bibfnamefont {V.~D.}\ \bibnamefont {Kurilovich}}, \bibinfo {author} {\bibfnamefont {J.}~\bibnamefont {Lebreuilly}},\ and\ \bibinfo {author} {\bibfnamefont {S.}~\bibnamefont {Girvin}},\ }\bibfield  {title} {\bibinfo {title} {Stabilizing the laughlin state of light: Dynamics of hole fractionalization},\ }\href {https://scipost.org/10.21468/SciPostPhys.13.5.107} {\bibfield  {journal} {\bibinfo  {journal} {SciPost Phys.}\ }\textbf {\bibinfo {volume} {13}} (\bibinfo {year} {2022})}\BibitemShut {NoStop}%
\bibitem [{\citenamefont {Shabani}\ and\ \citenamefont {Neven}(2016)}]{Shabani2016-go}%
  \BibitemOpen
  \bibfield  {author} {\bibinfo {author} {\bibfnamefont {A.}~\bibnamefont {Shabani}}\ and\ \bibinfo {author} {\bibfnamefont {H.}~\bibnamefont {Neven}},\ }\bibfield  {title} {\bibinfo {title} {Artificial quantum thermal bath: Engineering temperature for a many-body quantum system},\ }\href {http://dx.doi.org/10.1103/PhysRevA.94.052301} {\bibfield  {journal} {\bibinfo  {journal} {Phys. Rev. A (Coll. Park.)}\ }\textbf {\bibinfo {volume} {94}},\ \bibinfo {pages} {052301} (\bibinfo {year} {2016})}\BibitemShut {NoStop}%
\bibitem [{\citenamefont {Lewis-Swan}\ \emph {et~al.}(2019)\citenamefont {Lewis-Swan}, \citenamefont {Safavi-Naini}, \citenamefont {Kaufman},\ and\ \citenamefont {Rey}}]{Lewis-Swan2019-qd}%
  \BibitemOpen
  \bibfield  {author} {\bibinfo {author} {\bibfnamefont {R.~J.}\ \bibnamefont {Lewis-Swan}}, \bibinfo {author} {\bibfnamefont {A.}~\bibnamefont {Safavi-Naini}}, \bibinfo {author} {\bibfnamefont {A.~M.}\ \bibnamefont {Kaufman}},\ and\ \bibinfo {author} {\bibfnamefont {A.~M.}\ \bibnamefont {Rey}},\ }\bibfield  {title} {\bibinfo {title} {Dynamics of quantum information},\ }\href {https://www.nature.com/articles/s42254-019-0090-y} {\bibfield  {journal} {\bibinfo  {journal} {Nature Reviews Physics}\ }\textbf {\bibinfo {volume} {1}},\ \bibinfo {pages} {627} (\bibinfo {year} {2019})}\BibitemShut {NoStop}%
\bibitem [{\citenamefont {Yanay}\ and\ \citenamefont {Clerk}(2020)}]{Yanay2020-go}%
  \BibitemOpen
  \bibfield  {author} {\bibinfo {author} {\bibfnamefont {Y.}~\bibnamefont {Yanay}}\ and\ \bibinfo {author} {\bibfnamefont {A.~A.}\ \bibnamefont {Clerk}},\ }\bibfield  {title} {\bibinfo {title} {Reservoir engineering with localized dissipation: Dynamics and prethermalization},\ }\href {https://link.aps.org/doi/10.1103/PhysRevResearch.2.023177} {\bibfield  {journal} {\bibinfo  {journal} {Physical Review Research}\ }\textbf {\bibinfo {volume} {2}},\ \bibinfo {pages} {023177} (\bibinfo {year} {2020})}\BibitemShut {NoStop}%
\bibitem [{\citenamefont {Ares}\ \emph {et~al.}(2025)\citenamefont {Ares}, \citenamefont {Calabrese},\ and\ \citenamefont {Murciano}}]{Ares2025-qf}%
  \BibitemOpen
  \bibfield  {author} {\bibinfo {author} {\bibfnamefont {F.}~\bibnamefont {Ares}}, \bibinfo {author} {\bibfnamefont {P.}~\bibnamefont {Calabrese}},\ and\ \bibinfo {author} {\bibfnamefont {S.}~\bibnamefont {Murciano}},\ }\bibfield  {title} {\bibinfo {title} {The quantum mpemba effects},\ }\href {http://dx.doi.org/10.1038/s42254-025-00838-0} {\bibfield  {journal} {\bibinfo  {journal} {Nat. Rev. Phys.}\ }\textbf {\bibinfo {volume} {7}},\ \bibinfo {pages} {451} (\bibinfo {year} {2025})}\BibitemShut {NoStop}%
\bibitem [{\citenamefont {Campbell}\ \emph {et~al.}(2025)\citenamefont {Campbell}, \citenamefont {D'Amico}, \citenamefont {Ciampini}, \citenamefont {Anders}, \citenamefont {Ares}, \citenamefont {Artini}, \citenamefont {Auff\`{e}ves}, \citenamefont {Oftelie}, \citenamefont {Bettmann}, \citenamefont {Bonan\c{c}a} \emph {et~al.}}]{Campbell2025-il_truncate}%
  \BibitemOpen
  \bibfield  {author} {\bibinfo {author} {\bibfnamefont {S.}~\bibnamefont {Campbell}}, \bibinfo {author} {\bibfnamefont {I.}~\bibnamefont {D'Amico}}, \bibinfo {author} {\bibfnamefont {M.~A.}\ \bibnamefont {Ciampini}}, \bibinfo {author} {\bibfnamefont {J.}~\bibnamefont {Anders}}, \bibinfo {author} {\bibfnamefont {N.}~\bibnamefont {Ares}}, \bibinfo {author} {\bibfnamefont {S.}~\bibnamefont {Artini}}, \bibinfo {author} {\bibfnamefont {A.}~\bibnamefont {Auff\`{e}ves}}, \bibinfo {author} {\bibfnamefont {L.~B.}\ \bibnamefont {Oftelie}}, \bibinfo {author} {\bibfnamefont {L.~P.}\ \bibnamefont {Bettmann}}, \bibinfo {author} {\bibfnamefont {M.~V.~S.}\ \bibnamefont {Bonan\c{c}a}}, \emph {et~al.},\ }\bibfield  {title} {\bibinfo {title} {Roadmap on quantum thermodynamics},\ }\href {http://dx.doi.org/10.1088/2058-9565/ae1e27} {\bibfield  {journal} {\bibinfo  {journal} {Quantum Sci. Technol.}\ }\textbf {\bibinfo {volume} {11}},\ \bibinfo {pages} {012501} (\bibinfo {year} {2025})}\BibitemShut {NoStop}%
\bibitem [{\citenamefont {Aamir}\ \emph {et~al.}(2025)\citenamefont {Aamir}, \citenamefont {Jamet~Suria}, \citenamefont {Mar\'{\i}n~Guzm\'{a}n}, \citenamefont {Castillo-Moreno}, \citenamefont {Epstein}, \citenamefont {Yunger~Halpern},\ and\ \citenamefont {Gasparinetti}}]{Aamir2025-kz}%
  \BibitemOpen
  \bibfield  {author} {\bibinfo {author} {\bibfnamefont {M.~A.}\ \bibnamefont {Aamir}}, \bibinfo {author} {\bibfnamefont {P.}~\bibnamefont {Jamet~Suria}}, \bibinfo {author} {\bibfnamefont {J.~A.}\ \bibnamefont {Mar\'{\i}n~Guzm\'{a}n}}, \bibinfo {author} {\bibfnamefont {C.}~\bibnamefont {Castillo-Moreno}}, \bibinfo {author} {\bibfnamefont {J.~M.}\ \bibnamefont {Epstein}}, \bibinfo {author} {\bibfnamefont {N.}~\bibnamefont {Yunger~Halpern}},\ and\ \bibinfo {author} {\bibfnamefont {S.}~\bibnamefont {Gasparinetti}},\ }\bibfield  {title} {\bibinfo {title} {Thermally driven quantum refrigerator autonomously resets a superconducting qubit},\ }\href {http://dx.doi.org/10.1038/s41567-024-02708-5} {\bibfield  {journal} {\bibinfo  {journal} {Nat. Phys.}\ }\textbf {\bibinfo {volume} {21}},\ \bibinfo {pages} {318} (\bibinfo {year} {2025})}\BibitemShut {NoStop}%
\bibitem [{\citenamefont {Uusn{\"{a}}kki}\ \emph {et~al.}(2026)\citenamefont {Uusn{\"{a}}kki}, \citenamefont {M{\"{o}}rstedt}, \citenamefont {Teixeira}, \citenamefont {Rasola},\ and\ \citenamefont {M{\"{o}}tt{\"{o}}nen}}]{Uusnakki2026-ll}%
  \BibitemOpen
  \bibfield  {author} {\bibinfo {author} {\bibfnamefont {T.}~\bibnamefont {Uusn{\"{a}}kki}}, \bibinfo {author} {\bibfnamefont {T.}~\bibnamefont {M{\"{o}}rstedt}}, \bibinfo {author} {\bibfnamefont {W.}~\bibnamefont {Teixeira}}, \bibinfo {author} {\bibfnamefont {M.}~\bibnamefont {Rasola}},\ and\ \bibinfo {author} {\bibfnamefont {M.}~\bibnamefont {M{\"{o}}tt{\"{o}}nen}},\ }\bibfield  {title} {\bibinfo {title} {Initial demonstration of a quantum heat engine based on dissipation-engineered superconducting circuits},\ }\href {https://www.nature.com/articles/s41467-026-72651-x} {\bibfield  {journal} {\bibinfo  {journal} {Nat. Commun.}\ } (\bibinfo {year} {2026})}\BibitemShut {NoStop}%
\bibitem [{\citenamefont {Guo}\ \emph {et~al.}(2026)\citenamefont {Guo}, \citenamefont {Du},\ and\ \citenamefont {Ma}}]{guo_2026_20190733}%
  \BibitemOpen
  \bibfield  {author} {\bibinfo {author} {\bibfnamefont {Q.}~\bibnamefont {Guo}}, \bibinfo {author} {\bibfnamefont {B.}~\bibnamefont {Du}},\ and\ \bibinfo {author} {\bibfnamefont {R.}~\bibnamefont {Ma}},\ }\bibfield  {title} {\bibinfo {title} {Data and simulation code for ``{Entangling Superconducting Qubits via Energy-Selective Local Reservoirs}''},\ }\href {https://doi.org/10.5281/zenodo.20190733} {10.5281/zenodo.20190733} (\bibinfo {year} {2026})\BibitemShut {NoStop}%
\bibitem [{\citenamefont {Elder}\ \emph {et~al.}(2020)\citenamefont {Elder}, \citenamefont {Wang}, \citenamefont {Reinhold}, \citenamefont {Hann}, \citenamefont {Chou}, \citenamefont {Lester}, \citenamefont {Rosenblum}, \citenamefont {Frunzio}, \citenamefont {Jiang},\ and\ \citenamefont {Schoelkopf}}]{Elder2020-ya}%
  \BibitemOpen
  \bibfield  {author} {\bibinfo {author} {\bibfnamefont {S.~S.}\ \bibnamefont {Elder}}, \bibinfo {author} {\bibfnamefont {C.~S.}\ \bibnamefont {Wang}}, \bibinfo {author} {\bibfnamefont {P.}~\bibnamefont {Reinhold}}, \bibinfo {author} {\bibfnamefont {C.~T.}\ \bibnamefont {Hann}}, \bibinfo {author} {\bibfnamefont {K.~S.}\ \bibnamefont {Chou}}, \bibinfo {author} {\bibfnamefont {B.~J.}\ \bibnamefont {Lester}}, \bibinfo {author} {\bibfnamefont {S.}~\bibnamefont {Rosenblum}}, \bibinfo {author} {\bibfnamefont {L.}~\bibnamefont {Frunzio}}, \bibinfo {author} {\bibfnamefont {L.}~\bibnamefont {Jiang}},\ and\ \bibinfo {author} {\bibfnamefont {R.~J.}\ \bibnamefont {Schoelkopf}},\ }\bibfield  {title} {\bibinfo {title} {High-fidelity measurement of qubits encoded in multilevel superconducting circuits},\ }\href {http://dx.doi.org/10.1103/PhysRevX.10.011001} {\bibfield  {journal} {\bibinfo  {journal} {Phys. Rev. X.}\ }\textbf {\bibinfo {volume} {10}},\ \bibinfo {pages} {011001} (\bibinfo {year} {2020})}\BibitemShut {NoStop}%
\bibitem [{\citenamefont {{X. Y. Jin, A. Kamal, A. P. Sears, T. Gudmundsen, D. Hover, J. Miloshi, R. Slattery, F. Yan, J. Yoder, T. P. Orlando, S. Gustavsson, and W. D. Oliver}}(2015)}]{jin2015thermal}%
  \BibitemOpen
  \bibfield  {author} {\bibinfo {author} {\bibnamefont {{X. Y. Jin, A. Kamal, A. P. Sears, T. Gudmundsen, D. Hover, J. Miloshi, R. Slattery, F. Yan, J. Yoder, T. P. Orlando, S. Gustavsson, and W. D. Oliver}}},\ }\bibfield  {title} {\bibinfo {title} {Thermal and residual excited-state population in a 3d transmon qubit},\ }\href {https://doi.org/10.1103/PhysRevLett.114.240501} {\bibfield  {journal} {\bibinfo  {journal} {Physical Review Letters}\ }\textbf {\bibinfo {volume} {114}},\ \bibinfo {pages} {240501} (\bibinfo {year} {2015})}\BibitemShut {NoStop}%
\bibitem [{\citenamefont {Du}\ \emph {et~al.}(2026)\citenamefont {Du}, \citenamefont {Guo},\ and\ \citenamefont {Ma}}]{du2026programmable}%
  \BibitemOpen
  \bibfield  {author} {\bibinfo {author} {\bibfnamefont {B.}~\bibnamefont {Du}}, \bibinfo {author} {\bibfnamefont {Q.}~\bibnamefont {Guo}},\ and\ \bibinfo {author} {\bibfnamefont {R.}~\bibnamefont {Ma}},\ }\bibfield  {title} {\bibinfo {title} {Programmable superradiance in an interacting qubit array},\ }\href@noop {} {\bibfield  {journal} {\bibinfo  {journal} {arXiv preprint arXiv:2605.12442}\ } (\bibinfo {year} {2026})}\BibitemShut {NoStop}%
\bibitem [{\citenamefont {Elben}\ and\ \citenamefont {Vermersch}(2026)}]{elben2025randommeasjl}%
  \BibitemOpen
  \bibfield  {author} {\bibinfo {author} {\bibfnamefont {A.}~\bibnamefont {Elben}}\ and\ \bibinfo {author} {\bibfnamefont {B.}~\bibnamefont {Vermersch}},\ }\bibfield  {title} {\bibinfo {title} {Randommeas. jl: A julia package for randomized measurements in quantum devices},\ }\href {https://quantum-journal.org/papers/q-2026-04-28-2086/} {\bibfield  {journal} {\bibinfo  {journal} {Quantum}\ }\textbf {\bibinfo {volume} {10}},\ \bibinfo {pages} {2086} (\bibinfo {year} {2026})}\BibitemShut {NoStop}%
\bibitem [{\citenamefont {Smolin}\ \emph {et~al.}(2012)\citenamefont {Smolin}, \citenamefont {Gambetta},\ and\ \citenamefont {Smith}}]{PhysRevLett.108.070502}%
  \BibitemOpen
  \bibfield  {author} {\bibinfo {author} {\bibfnamefont {J.~A.}\ \bibnamefont {Smolin}}, \bibinfo {author} {\bibfnamefont {J.~M.}\ \bibnamefont {Gambetta}},\ and\ \bibinfo {author} {\bibfnamefont {G.}~\bibnamefont {Smith}},\ }\bibfield  {title} {\bibinfo {title} {Efficient method for computing the maximum-likelihood quantum state from measurements with additive gaussian noise},\ }\href {https://doi.org/10.1103/PhysRevLett.108.070502} {\bibfield  {journal} {\bibinfo  {journal} {Phys. Rev. Lett.}\ }\textbf {\bibinfo {volume} {108}},\ \bibinfo {pages} {070502} (\bibinfo {year} {2012})}\BibitemShut {NoStop}%
\end{thebibliography}%

\end{document}